\documentclass[pdflatex,sn-nature,iicol]{sn-jnl}

\usepackage{graphicx}%
\usepackage{multirow}%
\usepackage{amsmath,amssymb,amsfonts}%
\usepackage{amsthm}%
\usepackage{mathrsfs}%
\usepackage[title]{appendix}%
\usepackage{xcolor}%
\usepackage{textcomp}%
\usepackage{manyfoot}%
\usepackage{booktabs}%
\usepackage{algorithm}%
\usepackage{algorithmicx}%
\usepackage{algpseudocode}%
\usepackage{listings}%
\usepackage{siunitx}
\usepackage{float}
\usepackage{soul} 

\usepackage{lineno}  

\theoremstyle{thmstyleone}%
\theoremstyle{thmstyletwo}%

\theoremstyle{thmstylethree}%

\begin{document}

\title[Article Title]{
First measurement of reactor neutrino oscillations at JUNO
}


\author[7]{Angel Abusleme}
\author[52]{Thomas Adam}
\author[58]{Kai Adamowicz}
\author[12]{David Adey}
\author[74]{Shakeel Ahmad}
\author[74]{Rizwan Ahmed}
\author[49]{Timo Ahola}
\author[64]{Sebastiano Aiello}
\author[23]{Fengpeng An}
\author[12]{Guangpeng An}
% ...


% ...
\affil[7]{Pontificia Universidad Cat\'{o}lica de Chile, Santiago, Chile}
% ...
\affil[12]{Institute of High Energy Physics, Beijing, China}
% ...
\affil[23]{Sun Yat-sen University, Guangzhou, China}
% ...
\affil[49]{University of Jyvaskyla, Department of Physics, Jyvaskyla, Finland}
% ...
\affil[52]{IPHC, Universit\'{e} de Strasbourg, CNRS/IN2P3, F-67037 Strasbourg, France}
% ...
\affil[58]{III. Physikalisches Institut B, RWTH Aachen University, Aachen, Germany}
% ...
\affil[64]{INFN Catania and Dipartimento di Fisica e Astronomia dell Universit\`{a} di Catania, Catania, Italy}
% ...
\affil[74]{Pakistan Institute of Nuclear Science and Technology, Islamabad, Pakistan}
% ...






\abstract{
Neutrino oscillations~\cite{Kajita:2016cak,McDonald:2016ixn}, a quantum effect manifesting at macroscopic scales, are governed by lepton flavor mixing angles and neutrino mass-squared differences~\cite{ParticleDataGroup:2024cfk} 
that are fundamental parameters of particle physics, representing phenomena beyond the Standard Model.
Precision measurements of these parameters are essential for testing the completeness of the three-flavor framework, determining the mass ordering of neutrinos, and probing possible new physics. 
The Jiangmen Underground Neutrino Observatory (JUNO)~\cite{JUNO:2022hxd} 
is a 20\,kton liquid-scintillator detector located 52.5\,km from multiple reactor cores, designed to resolve the interference pattern of reactor neutrinos with sub-percent precision~\cite{JUNO:2024jaw, JUNO:2022mxj}. 
Here we report, using the first 59.1\,days of data collected since detector completion in August 2025, the first simultaneous high-precision determination of two neutrino oscillation parameters, $\sin^2 \theta_{12} = 0.3092\,\pm\,0.0087$ and $\Delta m^2_{21} = (7.50\,\pm\,0.12)\times10^{-5}\;{\rm eV}^2$ for the normal mass ordering scenario, improving the precision by a factor of 1.6 relative to the combination of all previous measurements.
These results advance the basic understanding of neutrinos, validate the detector’s design, and confirm JUNO’s readiness for its primary goal of resolving the neutrino mass ordering with a larger dataset.
The rapid achievement with a short exposure highlights JUNO’s potential to push the frontiers of precision neutrino physics and paves the way for its broad scientific program.}




\maketitle

Neutrinos and antineutrinos are fundamental particles that interact via the weak force in the electron ($\nu_e$/$\bar{\nu}_e$), muon ($\nu_\mu$/$\bar{\nu}_{\mu}$), and tau ($\nu_\tau$/$\bar{\nu}_{\tau}$) flavours.
They propagate as coherent superpositions of the mass eigenstates $\nu_1$, $\nu_2$, and $\nu_3$ with corresponding masses $m_1$, $m_2$, and $m_3$, allowing them to transform from one flavor to another.
 This phenomenon, known as neutrino oscillation~\cite{Kajita:2016cak,McDonald:2016ixn} and arising from the mismatch between the flavor and mass eigenstates, is described by three mixing angles ($\theta_{12}$, $\theta_{23}$, $\theta_{13}$), two independent mass-squared differences ($\Delta m^2_{21}\equiv m^2_2 - m^2_1$, $\Delta m^2_{31}\equiv m^2_3 - m^2_1$), and one CP-violating phase.
Precision determination of these parameters provides a stringent test of the unitarity of lepton flavor mixing~\cite{Pontecorvo:1967fh,Maki:1962mu}, validates the three-flavour framework, and constrains explorations of the Majorana nature of neutrinos, the matter-antimatter asymmetry, and the large-scale structure of the universe~\cite{Xing:2011zza,ParticleDataGroup:2024cfk}.

A global programme using solar, atmospheric, reactor, and accelerator neutrinos has yielded increasingly precise determinations of neutrino oscillation parameters. Solar neutrino experiments~\cite{Super-Kamiokande:2001ljr,SNO:2002tuh} first measured the parameters $\theta_{12}$ and $\Delta m^2_{21}$, hereafter referred to as the ``solar parameters''. Reactor experiments later confirmed this oscillation pattern~\cite{KamLAND:2002uet} and provided precise measurements of $\Delta m^2_{21}$~\cite{KamLAND:2013rgu}.
Atmospheric~\cite{Super-Kamiokande:1998kpq} and accelerator~\cite{K2K:2002icj} experiments determined $|\Delta m^2_{31}|$ and $\theta_{23}$, referred to as the ``atmospheric parameters''.
Short-baseline reactor experiments established a nonzero $\theta_{13}$~\cite{DayaBay:2012fng}, linking the solar and atmospheric sectors.

The JUNO experiment was proposed to determine which Neutrino Mass Ordering (NMO), normal ($m_1 < m_2 < m_3$) or inverted ($m_3 < m_1 < m_2$), exists in nature~\cite{Petcov:2001sy,Choubey:2003qx,Learned:2006wy,Li:2013zyd,Zhan:2008id,Zhan:2009rs}, by detecting electron antineutrinos produced in the $\beta$ decays of fission products from nuclear reactors. The JUNO site, located beneath Dashi Hill in Guangdong, China, is shielded by 650\,m of rock overburden ($\sim$1800\,m water equivalent) from cosmic radiation. The detector is positioned near the first maximum of the solar-driven oscillations, at an optimized distance of 52.5\,km, equidistant from all reactor cores of the Yangjiang and Taishan Nuclear Power Plants (NPP)~\cite{An:2015jdp,Li:2013zyd,JUNO:2024jaw}. At this baseline, the interference between oscillation frequencies produces a maximal spectral distortion in the reactor neutrino energy spectrum, from which the oscillation parameters can be precisely extracted. JUNO is a liquid scintillator (LS) detector that, thanks to advances in detector size~\cite{JUNO:2023ete} and energy resolution~\cite{JUNO:2024fdc}, inaugurates a new era of precision in the neutrino sector, with projected sub-percent sensitivity to $\Delta m_{21}^2$, $\Delta m_{31}^2$, and $\sin^2\theta_{12}$~\cite{JUNO:2022mxj}.
JUNO also has vast potential~\cite{An:2015jdp,JUNO:2022hxd} in astroparticle physics~\cite{JUNO:2023dnp,JUNO:2022lpc,JUNO:2025sfc,JUNO:2021tll,JUNO:2023zty,JUNO:2020hqc,JUNO:2022jkf} and in searches beyond the Standard Model~\cite{JUNO:2022qgr,Wang:2020uvi,Liu:2023cqs,JUNO:2023vyz,JUNO:2024pur}.

This paper presents JUNO’s first physics results, marking the end of a decade-long effort from the start of construction in 2015 to detector completion in August 2025. Based on 59.1 days of initial data (corresponding to 69 calendar days), our findings showcase JUNO’s performance and yield the most precise and simultaneous measurement of solar oscillation parameters $\Delta m^2_{21}$ and $\sin^2\theta_{12}$ to date. They also establish the foundation for JUNO’s long-term program to determine the NMO, test the completeness of the three-flavor framework, and probe possible new physics.

\begin{figure}[t]
\centering
\includegraphics[width=0.5\textwidth, angle = 0]
{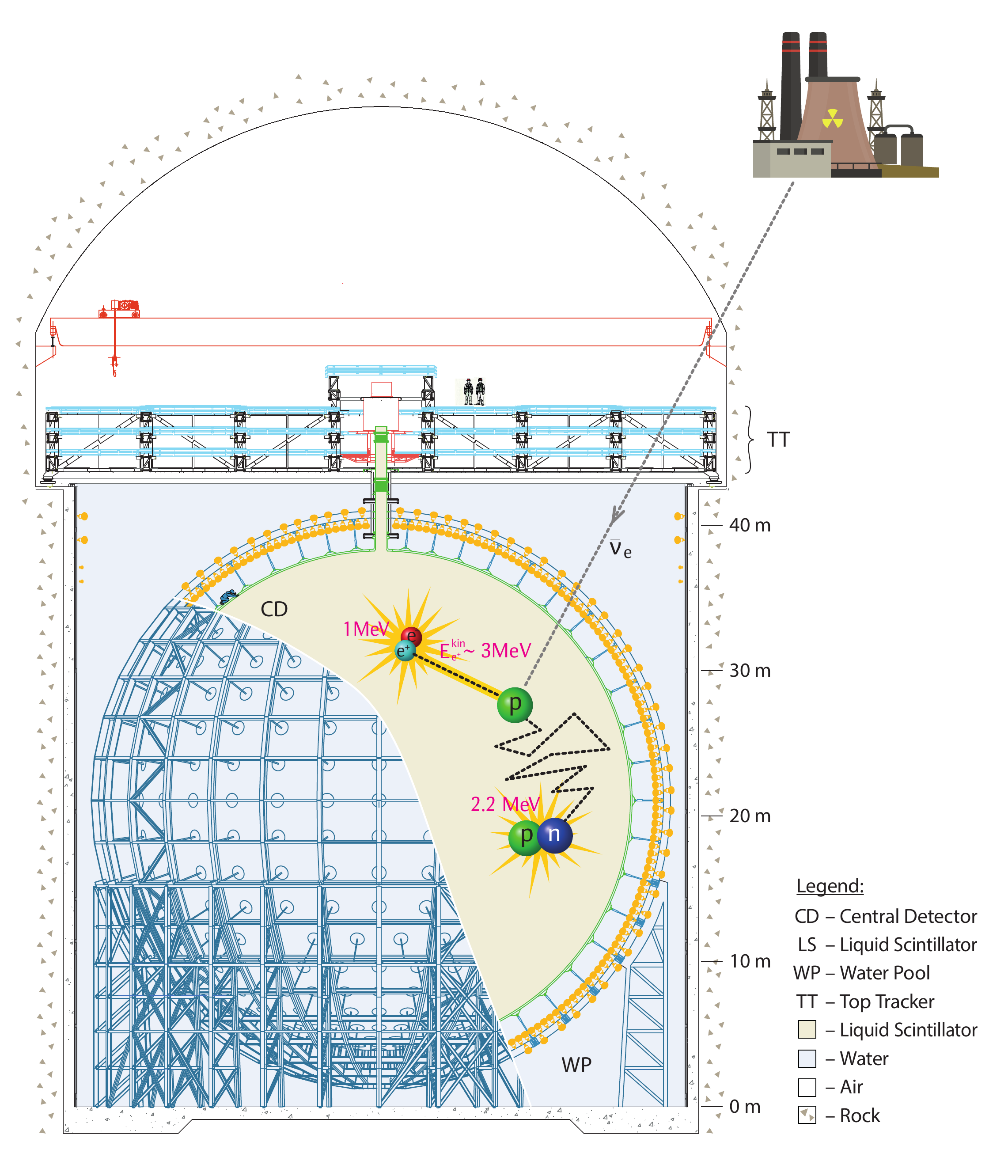}
\caption{{\bf JUNO experimental layout.} The 20\,kton liquid scintillator detector is located underground to detect electron antineutrinos from nuclear reactors through inverse beta decay interaction, cartooned inside the detector.}
\label{fig:JUNO}
\end{figure}

\section * {JUNO experimental setup}

JUNO comprises of a LS Central Detector (CD), surrounded by a large Water Pool (WP), and an external plastic-scintillator Top Tracker (TT), as illustrated in Fig.~\ref{fig:JUNO}.
The 20-kton LS target is contained within a 17.7-m-radius Acrylic Vessel (AV), viewed by 17,596 20-inch (4939 dynode and 12657 Micro-Channel Plate (MCP)) and 25,587 3-inch photomultiplier tubes (PMTs)~\cite{JUNOComiss}, which record the intensity and timing of scintillation light with 78\% total geometrical coverage.
A high-purity water buffer ($\ge$1.5\,m thickness) separates the vessel from the PMTs, while an outer Water Cherenkov Detector (WCD), at least 1.55\,m thick, shields against radioactivity and tags cosmic muons and their subsequent cosmogenic backgrounds. 
The TT partially covers the CD to provide a high-purity and high-precision muon sample for evaluating muon tagging efficiency and tracking precision in the CD.

To resolve the fine oscillation structure necessary for determining the NMO, JUNO is designed to achieve an energy resolution of 3\% at 1\,MeV~\cite{JUNO:2024fdc}, unprecedented for a detector of this scale. 
This performance requires a correspondingly high light yield so that statistical fluctuations remain sufficiently small.
Energy-scale non-linearity, mainly from ionization quenching and Cherenkov light emission, is controlled to sub-percent precision through an extensive calibration program using radioactive sources, a laser system, and two different PMT systems~\cite{JUNO:2020xtj}.

Further details on the JUNO detector design and instrumentation are provided in Methods and in Ref.~\cite{JUNOComiss} describing the initial performance results of the JUNO detector.

\section * {Data analysis}
\label{sec:neuAna}

Reactor neutrinos with energies up to about 11\,MeV are identified via inverse beta decay (IBD), $\overline{\nu}_e + p \rightarrow e^{+} + n$, where the prompt positron and delayed neutron capture ($\tau \sim 200$\,$\mu$s) provide distinctive signatures in space, time, and energy (Fig.~\ref{fig:JUNO}). The prompt energy relates to the antineutrino energy through $E_{p} \approx E_{\bar{\nu}_e} - 0.784$\,MeV. Given the 1.8\,MeV IBD kinematic threshold~\cite{Vogel:1999zy,Strumia:2003zx}, the observable $E_{p}$ spectrum begins at approximately 1\,MeV. Oscillation parameters are extracted from characteristic distortions in the $E_{p}$ spectrum via a multi-step analysis. Events are reconstructed and selected using criteria optimized for high signal efficiency and strong background suppression. The detector response, linking reconstructed to true energy, is calibrated using both deployed radioactive sources and naturally occurring backgrounds. Residual backgrounds are quantified using control samples or Monte Carlo (MC) simulations.

We used data from the 26$\mathrm{^{th}}$ of August to the 2$\mathrm{^{nd}}$ of November 2025 with a live-time of 59.1 days. Results, validated across three independent analyses, are consistent. 
One procedure is presented here, with methodological differences described in Methods.

{\bf Calibration and reconstruction:} The light yield, approximately 1785 photoelectrons (p.e.) per MeV, was calibrated using the neutron capture peak from an $^{241}$Am–$^{13}$C (AmC) source deployed at the detector center~\cite{JUNOComiss}. 
Event reconstruction uses the PMT waveform information, where the signal time represents the photon arrival time and the charge corresponds to the integrated signal in p.e.
Several likelihood algorithms combining these observables, with corrections accounting for spatial variations derived from the calibration data and natural sources, are employed to reconstruct event positions and energies, returning comparable performances. 

\begin{figure}[!h]
\centering
\includegraphics[width=0.48\textwidth]{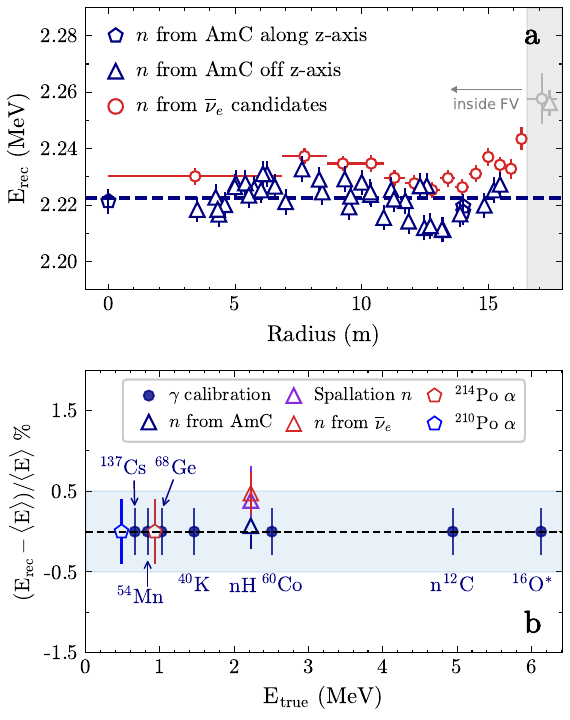}
\caption{
{\bf JUNO energy scale uncertainty.}
{\bf a}, Reconstructed energy $E_{\mathrm{rec}}$ of the neutron-capture peak ($E_\gamma=2.223$\,MeV, dashed line) from AmC neutron sources and antineutrino candidates across the full detector volume. The gray markers indicate neutrons outside the fiducial volume (FV). {\bf b}, Residual energy bias and systematic uncertainty for $\gamma$ sources at fixed calibration positions, $\alpha$-decays from natural radioactivity, and neutrons from muon-induced spallation and antineutrino candidates. $\langle E \rangle$ refers to expected reconstructed energy of each source at the center. Error bars includes residual temporal- and spatial-variations, and the band represents the estimated overall systematic uncertainty of the energy scale.}
\label{fig:energyUncer}
\end{figure}

The detector's energy response was characterized using multiple calibration sources, cosmogenic products, $^{214}$Po/$^{210}$Po, and neutrons from antineutrino candidates. Beyond periodic AmC calibrations, high-statistics spallation neutron events enabled daily-level corrections for energy scale.  Together with $^{214}$Po and $^{210}$Po samples, a residual temporal variation of 0.2\% was determined. The residual detector non-uniformity was found to be within 0.4\% in the fiducial volume (described in detail later) across the full dataset using $^{210}$Po and spallation neutrons, as also demonstrated by the results from AmC sources and antineutrino candidates shown in Fig.~\ref{fig:energyUncer}a. An overall energy uncertainty of 0.5\% was achieved, as shown in Fig.~\ref{fig:energyUncer}b.

Energy scale non-linearity comprises of scintillator non-linearity and instrumental non-linearity, directly affects the reconstructed antineutrino energy, and therefore the determination of oscillation parameters, see Ref.~\cite{DayaBay:2019fje}. Scintillator non-linearity originates mainly from ionization quenching (Birks' law~\cite{Birks:1964zz}) and Cherenkov light-emission, while instrumental non-linearity is caused by residual non-linearity of PMT charge reconstruction.
The non-linearity models were constrained by dedicated calibrations with $\gamma$'s covering around 0.5–6\,MeV, complemented by continuous $\beta^-$ spectrum (up to $\sim$14\,MeV) and $\beta^+$ spectrum (up to $\sim$2\,MeV) from cosmogenic isotopes. 
This yielded a positron energy non-linearity known with a precision of 1\% (see Methods) applied in the oscillation fit.

{\bf Antineutrino event selection:} The IBD signature is identified through the spatio-temporal coincidence between a prompt signal (positron kinetic energy deposition and its annihilation with an electron into two 0.511\,MeV $\gamma$'s) and a delayed signal (neutron capture on hydrogen producing a 2.223\,MeV $\gamma$). This distinctive signature provides strong background rejection while maintaining high signal efficiency.

Reactor neutrino candidates were selected using four principal criteria: spatio-temporal coincidence, a fiducial-volume (FV) cut to suppress external backgrounds, a dedicated muon veto to reject cosmogenic backgrounds, and a multiplicity cut to reduce instrumental and multi-particle coincidence backgrounds. Muons were identified with high efficiency ($>$99.9\%) via charge thresholds in both the central detector and water pool. Spallation neutrons were tagged based on energy and timing relative to preceding muons.
A temporal veto following muons removed short-lived muon-induced backgrounds, while a combined spatio-temporal veto around identified spallation neutrons suppressed long-lived cosmogenic products.

The FV cut selects prompt candidate vertices within 16.5\,m of the detector center, reducing contributions from radioactivity at the detector edge and avoiding regions with complex optical behaviour. Regions nearest to the detector’s poles are also excluded due to the complex structure and degraded reconstruction performance. 
The FV cut efficiency was evaluated from geometrical volume fraction (80.6\%) and is validated with uniformly distributed prompt IBD candidates with energies above 3.5\,MeV, to exclude contamination from non-homogeneously distributed external backgrounds, and an independent cosmogenic $^{12}$B sample. The residual non-uniformity near the fiducial boundary, evaluated using three independent vertex-reconstruction algorithms, results in a relative uncertainty of 1.6\%. A separate cross-check based on calibration measurements yields a consistent error estimate. More details are provided in Methods.

Other selection efficiencies—including the muon veto, spallation-neutron veto, and multiplicity cut—were derived via livetime analysis using 10-cm cubes. Energy selection efficiencies were validated using calibration sources: the prompt-energy threshold with $^{68}$Ge source, and the delayed-energy window via the neutron-capture peak of AmC source. Temporal and spatial coincidence efficiencies were derived from neutron-capture time distribution and the vertex-separation distribution of the IBD candidates, respectively, and were also verified using AmC sources and MC simulations. The efficiency due to the quality cut aimed to remove PMT flashers (spontaneous light emission from PMTs) was evaluated based on physics data such as $^{214}$Bi-$^{214}$Po, IBD, and calibration sources, as well as validated by MC simulations.

\begin{figure}[t]
\centering
\includegraphics[width=0.5\textwidth]{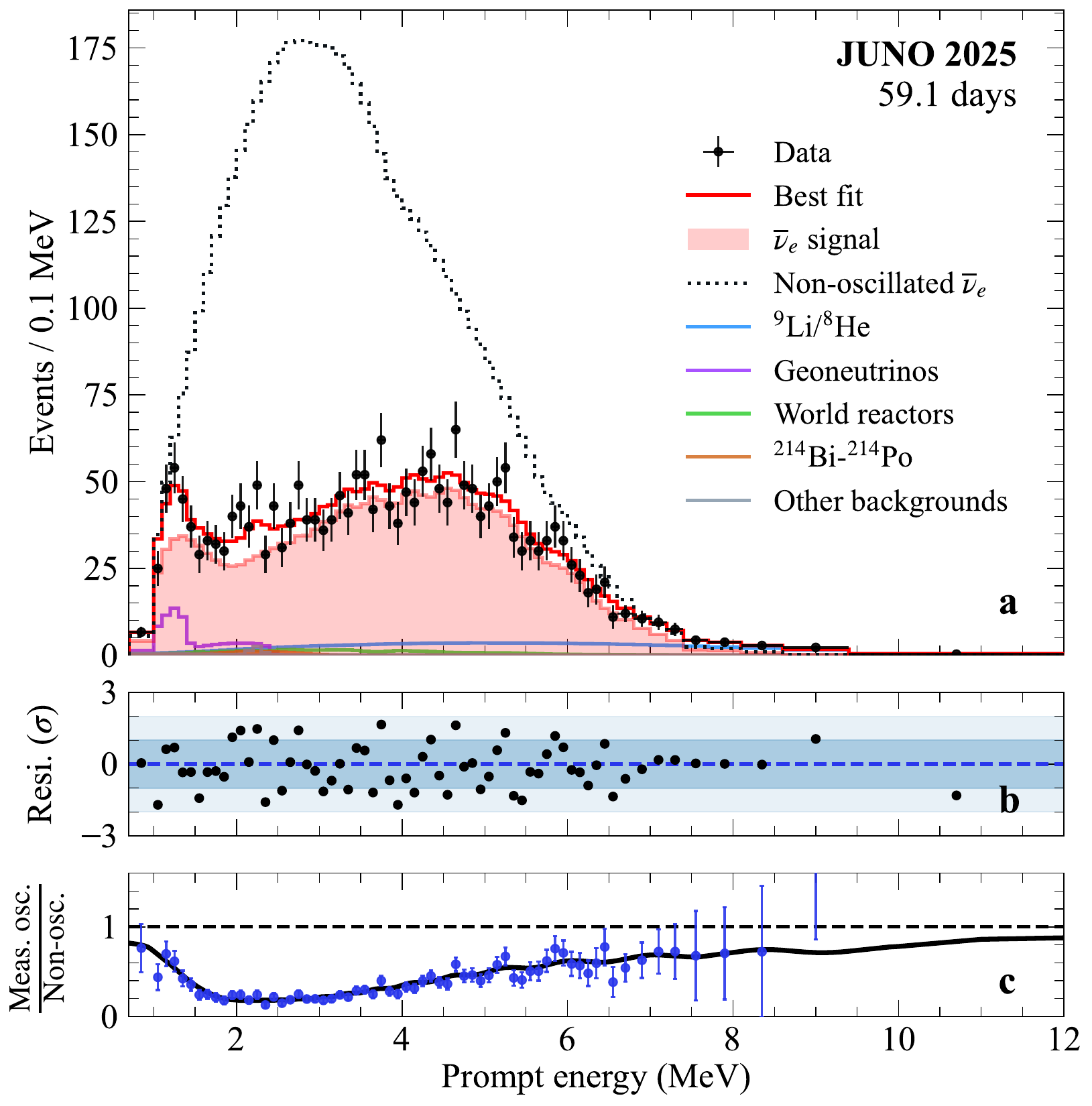}
\caption{{\bf Measured energy spectrum of prompt IBD candidates}. {\bf a}, Black points show measured data with statistical error bars, with the red curve indicating the best-fit oscillation model. Shaded red region represents expected antineutrino signal. Black dotted line represents non-oscillated reactor neutrino expectation. Backgrounds are indicated by other solid color lines. {\bf b}, Residuals quantifying the statistical consistency between data and the complete model. {\bf c}, Ratio of the measured oscillated spectrum to non-oscillated prediction.}
\label{fig:prompt_spectrum}
\end{figure}

The final data set contains 2379 identified antineutrino candidates. Their reconstructed prompt energy spectrum is presented in Fig. \ref{fig:prompt_spectrum}. 
All selection efficiencies and associated uncertainties are summarized in Table~\ref{tab:ibd}. 
The decomposed background components are also summarized in Table~\ref{tab:ibd}.

\begin{table}[!h]
\begin{tabular}{lcc}
\toprule
\multicolumn{3}{c}{\bf Antineutrinos ($\overline{\nu}_{e}$) Candidates Summary} \\
\midrule
DAQ live time (days) &  \multicolumn{2}{c}{59.1} \\
$\overline{\nu}_{e}$ candidates &  \multicolumn{2}{c}{2379} \\[0.5em]
{\bf Selection Efficiencies (\%)} & $\varepsilon$ & $\sigma_{\rm rel}$ \\
\quad Fiducial volume  &  80.6 & 1.6  \\
\quad PMT flasher rejection & $>$99.9 &  negligible \\
\quad $\mu$ veto   &  93.6  & negligible \\
\quad Multiplicity   &  97.4  & negligible \\
\quad Prompt-delayed coinc. & 95.1 & 0.13 \\
Total efficiency ($\varepsilon_{\rm tot}$) & 69.9 & 1.6 \\ [0.5em]
{\bf $\bf \overline{\nu}_{e}$ signal (cpd\footnotemark[1])} &  & \\ 
\quad w/o $\varepsilon_{\rm tot}$ corrected &  \multicolumn{2}{c}{33.5\,$\pm$\,1.7} \\
\quad w/{ } \,$\varepsilon_{\rm tot}$ corrected &  \multicolumn{2}{c}{47.9\,$\pm$\,2.6}  \\
\quad Non-oscillated $\overline{\nu}_{e}$ & \multicolumn{2}{c}{150.9\,$\pm$\,2.7} \\[0.5em]
{\bf  Backgrounds (cpd)} & Pre-fit & Best-fit \\ 
\quad $^9$Li/$^8$He  &  4.3\,$\pm$\,1.4 & 3.9\,$\pm$\,0.6  \\
\quad Geoneutrinos  &  1.2\,$\pm$\,0.5 & 1.4\,$\pm$\,0.4 \\
\quad World reactors &  0.88\,$\pm$\,0.09  & 0.88\,$\pm$\,0.09  \\
\quad $^{214}$Bi-$^{214}$Po &  0.18\,$\pm$\,0.10 & 0.20\,$\pm$\,0.10\\
\quad $^{13}$C($\alpha$, n)$^{16}$O  &  0.04\,$\pm$\,0.02  & 0.04\,$\pm$\,0.02 \\
\quad Fast neutrons  &  0.02\,$\pm$\,0.02 & 0.02\,$\pm$\,0.02  \\
\quad Double neutrons  &  0.05\,$\pm$\,0.05 & 0.07\,$\pm$\,0.05 \\
\quad Atmospheric neutrinos  &  0.08\,$\pm$\,0.04 & 0.07\,$\pm$\,0.04 \\
\quad Accidentals ($\times10^{-2}$) &  4.9\,$\pm$\,0.3 & 4.9\,$\pm$\,0.3\\
\botrule
\end{tabular}
\caption{
{\bf Event selection and backgrounds.} 
The table lists the IBD selection criteria with their respective efficiencies ($\varepsilon$) and relative uncertainties ($\sigma_{\rm rel}$), alongside the measured or expected pre-fit and best-fit background rates.
The $\overline{\nu}_e$ signal rates were derived both with and without applying the efficiency correction to the observed IBD rate, after background subtraction. The non-oscillated prediction is based on reactor operational data and the flux model described in Methods.
The best-fit values for the backgrounds come from the spectral fit.}
\footnotetext[1]{cpd $\equiv$ counts per day
}
\label{tab:ibd}
\end{table}

{\bf Background evaluation:} The main backgrounds for this analysis are represented by cosmogenic $^9$Li/$^8$He isotopes, geoneutrinos, $^{214}$Bi-$^{214}$Po, and antineutrinos from other world reactors. Other minor backgrounds are either strongly suppressed by the selection cuts, or are intrinsically small. More details are provided in Methods.

The dominant background originated from cosmogenic $^9$Li/$^8$He $\beta$-n decays was suppressed using a spallation neutron veto. The total and residual rates were evaluated by fitting the time distribution of candidate events since cosmic muons, before and after application of the spallation neutron veto, respectively. The $^9$Li/$^8$He ratio was constrained by the measurement from Daya Bay~\cite{DayaBay:2024xye}. The dependence of the $^9$Li/$^8$He rate on visible muon energy was compared with Geant4~\cite{GEANT4:2002zbu} simulations, and a systematic rate uncertainty of 34\% was assigned to encompass fitting uncertainties and data–simulation discrepancies. A high-purity sample of spallation-neutron-tagged $^9$Li/$^8$He events enabled spectrum comparison with both calculations based on nuclear databases and a parameterized model from Daya Bay data~\cite{DayaBay:2022orm}. This comparison constrained the spectral shape uncertainty to 20\%. Other long-lived $\beta$–n emitters were estimated using FLUKA~\cite{Battistoni:2015epi,Ahdida:2022gjl} and Geant4~\cite{GEANT4:2002zbu} simulations and found to be negligible.

Geoneutrinos, electron antineutrinos emitted in $^{238}$U and $^{232}$Th decay chains in the Earth's interior, are indistinguishable from reactor neutrinos on an event-by-event basis. Based on geological crust and mantle models, their predicted flux yields (1.1–2.4) IBD events per day interacting in the LS assuming 100\% detection efficiency~\cite{JUNO:2025sfc}. Taking the variations across the models, we adopted 1.75 events/day with a 40\% uncertainty. A 5\% shape uncertainty accounts for the two spectral models used~\cite{Enomoto,Li:2024mqz}, the $^{238}$U/$^{232}$Th compositional variations, and oscillation effects. An additional 10\% uncertainty was applied to the nominal $^{232}$Th-to-$^{238}$U signal ratio of 0.29.

World reactor backgrounds—antineutrinos from distant reactors beyond the primary signal sources—were estimated using operational data from Chinese nuclear plants~\cite{CNEA} and the global PRIS database~\cite{PRIS}. This background contributes to 1.26 IBD events per day at the JUNO site, with 10\% rate and 5\% shape uncertainties.

The $\beta$-$\alpha$ cascades from $^{214}$Bi-$^{214}$Po (radon daughters) introduce a correlated background, in which delayed $\alpha$ signals can have reconstructed energies elevated into the neutron capture region, either due to $\gamma$'s from excited $^{214}$Po states~\cite{Borexino:2019gps} or to elastic scattering on protons~\cite{SNO:2025koj}. Other backgrounds (Table~\ref{tab:ibd}) were measured or simulated and found to be subdominant (see Methods).

{\bf Reactor--neutrino signal prediction:} Antineutrinos in reactors are produced from $\beta$ decays of neutron-rich fission fragments of $^{235}$U, $^{238}$U, $^{239}$Pu, and $^{241}$Pu. The reactor operational data, especially thermal power, fission fractions and spent fuel information served as inputs to the flux prediction model. The flux model was decomposed into a component common to JUNO and the Daya Bay experiment~\cite{DayaBay:2021dqj,DayaBay:2025ngb}, plus a subdominant term covering their differences, including contributions from Taishan, Yangjiang, and other distant reactors. Minor corrections from spent nuclear fuel and non-equilibrium isotopes were also included. More details are given in Methods.

\section * {Spectral fit and results}
\label{sec:result}

 The observed prompt energy spectrum of IBD candidates, shown in Fig.~\ref{fig:prompt_spectrum}, was compared with a detailed prediction to determine the neutrino oscillation parameters. This prediction incorporated the complete sequence of effects from reactor neutrino production to signal detection (details in Methods): reactor fluxes evaluated via a semi-model-independent combination of Daya Bay near detector data~\cite{DayaBay:2021dqj, DayaBay:2025ngb} and theoretical models~\cite{Huber:2011wv, Mueller:2011nm}, baselines from JUNO site to multiple reactors as provided in Ref.~\cite{JUNO:2024jaw}, 
 time-dependent fission fractions, neutrino oscillation including Earth's matter effects~\cite{Wolfenstein:1977ue,Mikheyev:1985zog}, interaction cross section~\cite{Vogel:1999zy,Strumia:2003zx}, detector response, and estimated backgrounds (Table~\ref{tab:ibd}).

Systematic uncertainties in the reactor flux prediction, shown in Table~\ref{tab:Uncer},
are dominated by the 1.2\% contribution from the Daya Bay reference spectrum~\cite{DayaBay:2025ngb}, followed by the 1\% uncertainty in the JUNO target proton number, which is driven by the liquid scintillator volume and hydrogen fraction.

\begin{figure}[t]
\centering
\includegraphics[width=0.5\textwidth]{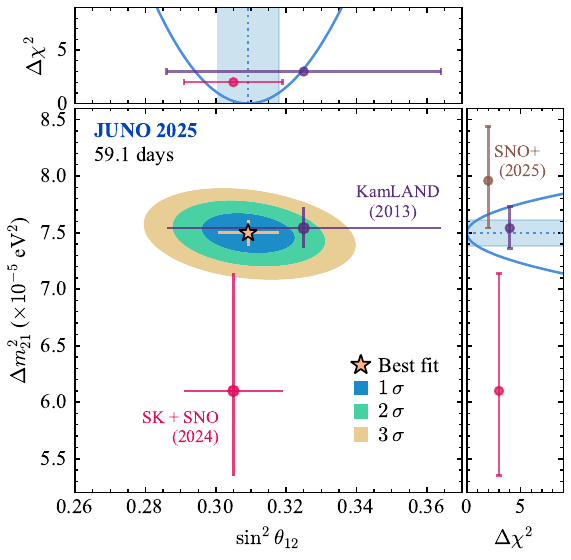}
\caption{{\bf Results on solar oscillation parameters.} Confidence intervals of $\sin^2\theta_{12}$ and $\Delta m^{2}_{21}$ from a spectral fit of measured prompt antineutrino energy spectrum shown in Fig.~\ref{fig:prompt_spectrum}. 
Shaded elliptically-shaped areas correspond to 1$\sigma$, 2$\sigma$, and 3$\sigma$ confidence levels. 
The upper panel provides the one dimensional $\Delta\chi^2$ for $\sin^2\theta_{12}$ obtained by profiling $\Delta m^{2}_{21}$ (blue line) and the blue shaded region as the corresponding 1$\sigma$ interval. The right panel is the same, but for $\Delta m^{2}_{21}$, with $\sin^2\theta_{12}$ profiled.
The star marks the best fit values of JUNO, and the error bars show their one-dimensional 1$\sigma$ confidence intervals. 
Results from other measurements of reactor neutrinos 
(KamLAND~\cite{KamLAND:2013rgu} and SNO+~\cite{SNO:2025koj}) and solar neutrinos (combined Super-Kamiokande (SK)+SNO~\cite{Super-Kamiokande:2023jbt}) are shown for comparison.}
\label{fig:contour}
\end{figure}

The spectral distortion observed in the prompt-energy spectrum in Fig.~\ref{fig:prompt_spectrum} was primarily shaped by solar oscillation parameters $\Delta m^2_{21}$ and $\sin^2\theta_{12}$, with a weaker dependence on $\Delta m^2_{31}$. Due to the currently limited statistics, our oscillation analysis concentrated on determining the solar parameters.
We derived the best-fit values and allowed regions using a $\chi^2$ statistic based on the Combined Neyman–Pearson (CNP) method~\cite{Ji:2019yca}, which incorporates statistical and systematic uncertainties of both signal and backgrounds as well as external constraints on parameters such as $\Delta m^2_{31}$ and $\sin^2\theta_{13}~$\cite{DayaBay:2022orm}. Leaving $\Delta m^2_{31}$ unconstrained had a negligible impact on the results. 
The variations in the Earth's crust density along the neutrino propagation trajectory were negligible, validating the use of a constant density ($\rho = (2.55 \pm 0.25)\ \mathrm{g\,cm^{-3}}$) approximation.
The robustness of the results was confirmed through independent analyses employing different fitting strategies. A blinding strategy was used to conceal the oscillation parameters throughout the analysis.  
In particular, reactor operational data, including power and fission fractions, were unblinded only after finalizing the analysis framework.

The oscillation analysis yielded the following results for the solar parameters for normal mass ordering scenario:
\begin{align}
\sin^2 \theta_{12} &= 0.3092\,\pm\,0.0087\,, \nonumber\\
\Delta m^2_{21} &= (7.50\,\pm\,0.12) \times 10^{-5}\; {\rm eV}^2\,.\nonumber
\end{align}
The results obtained for the inverted mass ordering scenario are fully compatible.
The allowed regions of oscillation parameters are shown in Fig.~\ref{fig:contour}. 
Consistent with the combination of all previous measurements~\cite{KamLAND:2013rgu,Super-Kamiokande:2023jbt}, this result improved the precision of both parameters by a factor of 1.6.

\section * {Discussion and outlook}
\label{sec:discussion}

JUNO achieves world-leading precision on the solar oscillation parameters, reducing the uncertainty in $\Delta m^{2}_{21}$ to 1.55\% (from KamLAND’s 2.5\%~\cite{KamLAND:2013rgu}) and in $\sin^2\theta_{12}$ to 2.81\% (from Super-Kamiokande+SNO’s 4.6\%~\cite{Super-Kamiokande:2023jbt}). 
 
The improved precision also supplies essential input to global studies of leptonic CP violation and neutrinoless double-beta decay. The current $\Delta m^2_{31}$ precision is limited by statistics in the 59.1-day dataset and parameter degeneracies. This precision will improve with increased statistics combined with a model-independent analysis using data from the JUNO's satellite experiment TAO~\cite{JUNO:2020ijm}.
Overall results confirm that the detector performance, calibration strategy, and analysis pipeline perform fully according to design expectations, providing the first quantitative validation of JUNO's capability for sub-percent precision oscillation measurements and future NMO determination.

\section*{Acknowledgment}
We gratefully acknowledge the continued cooperation and support of the China General Nuclear Power Group in the construction and operation of the JUNO experiment.
We acknowledge financial and institutional support from the Chinese Academy of Sciences, the National Key R\&D Program of China, the People's Government of Guangdong Province, and the Tsung--Dao Lee Institute of Shanghai Jiao Tong University in China.
We appreciate the contributions from the Institut National de Physique Nucl\'eaire et de Physique des Particules (IN2P3) in France, the Istituto Nazionale di Fisica Nucleare (INFN) in Italy, the Fonds de la Recherche Scientifique (F.R.S.--FNRS) and the Institut Interuniversitaire des Sciences Nucl\'eaires (IISN) in Belgium, and the Conselho Nacional de Desenvolvimento Cient\'ifico e Tecnol\'ogico (CNPq) in Brazil.
We also acknowledge the support of the Agencia Nacional de Investigaci\'on y Desarrollo (ANID) and the ANID--Millennium Science Initiative Program (ICN2019044) in Chile; the European Structural and Investment Funds, the Ministry of Education, Youth and Sports, and the Charles University Research Center in the Czech Republic; the Deutsche Forschungsgemeinschaft (DFG), the Helmholtz Association, and the Cluster of Excellence PRISMA+ in Germany; and the Joint Institute for Nuclear Research (JINR) and Lomonosov Moscow State University in Russia.
We further thank the Slovak Research and Development Agency in the Slovak Republic, the National Science and Technology Council (NSCT) and MOE in Taiwan, China, the Program Management Unit for Human Resources \& Institutional Development, Research and Innovation (PMU--B), Chulalongkorn University, and Suranaree University of Technology in Thailand, the Science and Technology Facilities Council (STFC) in the United Kingdom, and the University of California at Irvine and the National Science Foundation (NSF) in the United States.
We also acknowledge the computing resources provided by the Chinese Academy of Sciences, IN2P3, INFN, and JINR, which are essential for data processing and analysis within the JUNO Collaboration.

\section*{Author contributions}

The JUNO detector was designed, constructed, commissioned and is being operated by the JUNO Collaboration. All aspects of the hardware and software developments, data taking, detector calibration, data processing, Monte Carlo simulation, as well as data analysis were performed by JUNO members, who also discussed and approved scientific results. All authors reviewed and approved the final version of the manuscript.

\section * {Methods}
\label{sec:Methods}

\subsection * {Oscillation of reactor $\bar{\nu}_e$}
\label{subsec:3flavorOsci}

In the three-neutrino oscillation framework, the flavor states ($\nu_e$/$\bar{\nu}_e, \nu_\mu$/$\bar{\nu}_{\mu}, \nu_\tau$/$\bar{\nu}_{\tau}$) are connected with the mass eigenstates ($\nu_1$/$\bar{\nu}_1$, $\nu_2$/$\bar{\nu}_3$, $\nu_3$/$\bar{\nu}_3$) by the leptonic mixing matrix~\cite{Pontecorvo:1967fh,Maki:1962mu} parametrised by three mixing angles $\theta_{12}$, $\theta_{23}$, $\theta_{13}$ and one complex phase $\delta$~\cite{ParticleDataGroup:2024cfk}. The oscillation frequencies are determined by the three mass-squared differences
$\Delta m^2_{ij} \equiv m_i^2 - m_j^2$, $(i,j = 1,2,3,\, i>j)$,
where $m_i$ is the $\nu_i$ mass. 
Typical approximate values for the mixing angles are
$\theta_{12} \approx 33^\circ$, 
$\theta_{23} \approx 45^\circ$, and 
$\theta_{13} \approx 8.5^\circ$.
The solar mass-squared splitting is 
$\Delta m^2_{21} \simeq 7.4\times10^{-5}\,\text{eV}^2$, 
while the atmospheric splitting is about two orders of magnitude larger,
$|\Delta m^2_{31}| \simeq 2.5\times10^{-3}\,\text{eV}^2$.
While ordering of the first two mass states ($m_2 > m_1$) is known, 
the ordering of $m_3$---normal or inverted---remains undetermined.

When neutrinos propagate in matter, oscillation parameters acquire the corrections from the effective matter potential~\cite{Wolfenstein:1977ue,Mikheyev:1985zog}, denoted by tilded quantities $\widetilde\theta_{ij}$ and $\Delta \widetilde m^2_{ij}$. For JUNO, matter effects are minor~\cite{Li:2016txk} and its main sensitivity to oscillation parameters  $\sin^2\theta_{12}$, $\Delta m^2_{21}$, and $\Delta m^2_{31}$ and to  NMO arises from vacuum oscillations.

\begin{figure}[!h]
\centering
\includegraphics[width=0.5\textwidth]{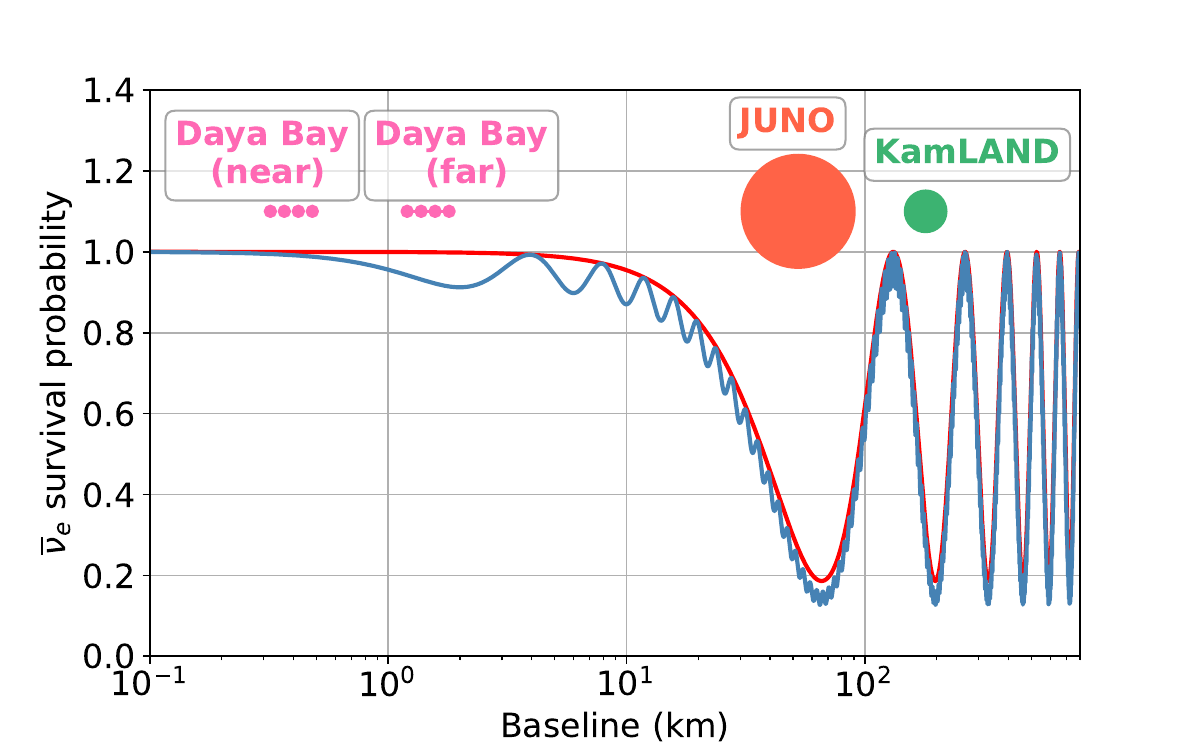}
\caption{{\bf Reactor neutrino oscillation at different baselines.} Survival probability ${\cal P}_{ee}$ of 4\,MeV reactor $\bar{\nu}_e$ vs. baseline. Red: slow solar oscillation ($\Delta m^2_{21}$); blue: total ${\cal P}{ee}$ including fast atmospheric oscillations ($\Delta m^2_{31}$). Experimental baselines: JUNO (orange, 52.5\,km) near solar oscillation maximum, Daya Bay near/far (pink), KamLAND (green). Circle sizes scale with detector sizes.}
\label{fig:Pee_vs_L}
\end{figure}

The IBD interaction used to detect reactor neutrinos is sensitive only to the electron flavor; consequently, the relevant oscillation observable is the electron antineutrino survival probability ${\cal P}_{ee} \equiv {\cal P} (\overline{\nu}_e \to \overline{\nu}_e)$ that can be expressed as:
\begin{eqnarray}
& \nonumber {\cal P}_{ee} = 1 - \sin^2 2\widetilde\theta_{12}\, \tilde c_{13}^4\, \sin^2 \widetilde\Delta_{21} \nonumber \\
& \quad - \sin^2 2\widetilde\theta_{13} \left( \tilde c_{12}^2 \sin^2 \widetilde\Delta_{31} + \tilde s_{12}^2 \sin^2 \widetilde\Delta_{32} \right), 
\label{eq:P_ee_aligned2}
\end{eqnarray}
where $\tilde c_{ij} \equiv \cos\widetilde\theta_{ij}$, $\tilde s_{ij} \equiv \sin\widetilde\theta_{ij}$, and $\widetilde\Delta_{ij} = \Delta \widetilde m^2_{ij} L / 4 E_{\bar\nu_e}$, with $E_{\bar\nu_e}$ being the antineutrino energy and $L$ the baseline from reactor cores to the detector.
The first ``solar term'' represents the slow oscillation associated with $\Delta m^2_{21}$, while the second ``atmospheric term'' characterizes the fast oscillation, as illustrated by the red and blue curves in Fig.~\ref{fig:Pee_vs_L}, respectively.

For a typical antineutrino energy of 4\,MeV, the solar oscillation length is $\sim$60\,km, while the atmospheric oscillation length is about 2\,km.
JUNO's 52.5\,km baseline is near the first solar oscillation maximum, allowing it to probe the interference with atmospheric oscillations and thus determine the NMO. 
Simultaneously, ${\cal P}_{ee}$ is near its minimum, forcing JUNO to be an exceptionally large detector to gather sufficient statistics for its physics goals.

\subsection* {JUNO detector setup}
\label{subsec:junoDetector}

The main JUNO detector components, CD, WP, and TT, are shown in Fig.~\ref{fig:JUNO}.

The CD~\cite{JUNO:2023ete} represents the core part of the experiment. It contains 20\,kton of LS enclosed in a 35.4\,m diameter spherical AV, constructed from highly transparent acrylic panels and held in place by a supporting Stainless Steel Structure (SSS). 
The optimization of the LS mixture, reported in Refs.~\cite{JUNO:2020bcl,Zhang:2020mqz}, was performed using a Daya Bay detector and a comprehensive optical model, with the results extrapolated to the JUNO scale.
The JUNO LS consists of linear alkyl benzene (LAB) as solvent, 2.5\,g/L PPO as primary fluor, 3\,mg/L bis-MSB as wavelength shifter, and 42.7~mg/L BHT as antioxidant to prevent long-term optical degradation. The 17,596 20-inch ``large'' photomultiplier tubes (LPMTs) provide the primary photon detection, achieving approximately 75\% geometrical coverage. This includes 4,939 dynode PMTs and 12,657 MCP PMTs mounted on the support structure. Additional 25,587 3-inch ``small'' PMTs (SPMTs) are installed among the LPMTs (3\% geometrical coverage) to extend the dynamic range and enhance calibration precision.

The CD is installed inside a cylindrical WP 43.5\,m in diameter and 44.0\,m in height, that is filled with 41\,kton of high-purity water ($^{238}$U/$^{232}$Th $< 10^{-15}$\,g/g, $^{222}\text{Rn}< 10$\,mBq/m$^3$, $^{226}\rm{Ra}< 50$\,$\mu$Bq/m$^3$) characterised by a light attenuation length of $> 60\,\text{m}$. The WP shields the CD against external radiation and acts as a WCD for the residual cosmic-ray muon flux. Additional 6\,kton of water are used in the CD region between the SSS and the AV as a shield against external background. 
 Cherenkov photons produced by cosmic muons are detected by 2752 MCP PMTs and additional 600 8-inch R5912 PMTs. The WCD is optically separated from the CD by Tyvek reflective foils. The muon tagging efficiency is very close to 100\%, with a very small geometrical inefficiency due to the top chimney, which connects the top of the CD to the calibration house above it.

The cosmic muon veto system is complemented by an external TT, consisting of three layers of plastic scintillator refurbished from the OPERA experiment~\cite{Acquafredda:2009zz}, covering approximately 60\% of the top surface above the WP. 
Its primary purpose is to provide a sample of muons, reconstructed with an angular precision of $0.20^\circ$~\cite{JUNO:2023cbw}, which can be used to benchmark the track reconstruction of the WP and CD.

The CD calibration strategy~\cite{JUNO:2020xtj} is designed to employ four source-deployment subsystems—the Automatic Calibration Unit (ACU), the Cable Loop System (CLS), the Guide Tube (GT), and a Remote Operated Vehicle (ROV)—which together allow precise placement of radioactive and optical sources along the detector’s central axis, within radial planes, at the detector boundary, and at arbitrary three-dimensional positions, respectively.
In the WCD, LED flashers installed along the WP wall and on the bottom are employed to perform the PMT timing calibration.

\subsection* {Detector filling and commissioning}
\label{subsec:junoDetector}

The JUNO detector was filled from December 2024 to August 2025 with the support of dedicated auxiliary facilities. 
The CD and WP were first filled simultaneously with water from the high-purity Water plants, then the water in the CD was gradually replaced with high-purity LS under nitrogen protection. The LS purification process included alumina filtering, vacuum distillation, mixing with PPO, bis-MSB, and BHT to form the master solution, water washing, and underground water extraction and nitrogen stripping. The Filling, Overflow, and Circulation system managed the liquid levels, ensured structural safety, and allowed online recirculation if needed. OSIRIS, a 20-ton scintillator detector~\cite{JUNO:2021wzm},  monitored LS samples for radioactivity during production. 

The detector commissioning was carried out simultaneously with the water filling and LS exchange.
%
%
JUNO online LPMT waveforms and event reconstructions enabled real-time $^{222}$Rn monitoring, keeping contamination below 1~mBq/m$^3$. The LS reached an attenuation length of 20.6\,m at 430\,nm, and radiopurity measurements in the FV showed $^{238}$U and $^{232}$Th levels of $(7.5\pm0.9)\times10^{-17}$\,g/g and $(8.2\pm0.7)\times10^{-17}$\,g/g, respectively, exceeding JUNO’s design requirements.
The background from unsupported $^{210}$Po, averaged over the analyzed period, was $(4.3 \pm 0.3) \times 10^{4}$ cpd/kton.
This performance results from stringent material selection, advanced purification, and meticulous assembly protocols- including continuous leak-checking, surface treatments, and clean-room installation~\cite{JUNO:2021kxb}.

JUNO uses full waveform readout of all LPMTs digitized by underwater Global Control Units and processed through a multiplicity-based global trigger. Each LPMT generates a ``hit'' when the signal exceeds five times the channel's noise. A global trigger is issued when 350 PMTs fire within a 304\,ns window, resulting in an overall trigger rate of $\sim$500\,Hz. The data acquisition gate has a duration of 1008\,ns. Instead, SPMTs have a self-trigger threshold of 0.33\,p.e. The Online Event Classification system performs real-time waveform and event reconstruction to select relevant physics events, for which full waveforms are stored, reducing the data rate to $\sim$90\,MB/s. No triggered events are discarded. 

For more information on the JUNO detector, see Ref.~\cite{JUNO:2022hxd}, while a detailed description of the commissioning and initial performance is available in Ref.~\cite{JUNOComiss}.

\subsection * {Calibration strategy} \label{subsec:CalibRec}

At the beginning of data taking, from August 23 to 26, a calibration campaign was carried out to calibrate the PMTs, establish the energy scale, and characterize the detector response.
The ACU deployed five $\gamma$ sources--$^{68}$Ge (2$\times$511 keV from $e^+$ annihilation), $^{137}$Cs (661.7 keV), $^{54}$Mn (834.8 keV), $^{40}$K (1460.8 keV), and $^{60}$Co (1173.2 keV and 1332.5 keV)--together with an AmC neutron source and a UV laser at the detector center and at two off-center positions ($\pm$14 m), with placement controlled to better than 1\,cm.
Additional AmC neutron-source calibration was performed: 7 points with the ACU and 39 off-axis points with the source attached to the CLS, enabling calibration along a vertical half-plane with a positioning precision of about 3\,cm.
 The light yield was biweekly monitored with the AmC source  deployed at the detector center.

The \textbf{LPMT calibration program} monitors charge response, gain, timing, and dark noise using weekly UV-laser runs and a 50\,Hz periodic trigger.
Laser LPMTs charge spectra were fit with models tailored separately for dynode PMTs and for MCP PMTs~\cite{JUNOComiss}.
The average mean gain—-$6.5\times10^{6}$ for dynode PMTs and $7.2\times10^{6}$ for MCP PMTs—remained stable over time within 0.2\%. %
The relative photon detection efficiency of each PMT was calibrated using isotropic light produced by AmC source at the CD center.
The Transit Time Spread (TTS) was calibrated with laser data from the full water phase.
The average values were 7.5 ns for MCP PMTs and 1.9 ns for dynode PMTs.

PMT timing calibration utilizes pulsed laser light delivered to the detector, with a fast monitor PMT providing a precise $t_0$ reference. Residual time distributions—obtained from first-hit times corrected for photon time-of-flight—allow the extraction of the TTS and the determination of per-channel time offsets. After calibration, the PMTs achieve sub-nanosecond synchronization across the whole detector.

PMT dark noise was monitored using periodic triggers, yielding average dark rates of 20.6 kHz for dynode PMTs and 22.7 kHz for MCP PMTs. Channels with low occupancy or timing misalignment were excluded from analysis.

\subsection * {Reconstruction} \label{subsec:CalibRec}

{\bf LPMT waveform reconstruction:} The JUNO analysis chain starts from waveform reconstruction of LPMTs.
The digitized waveforms are reconstructed to extract the pulse time ($t$) and charge ($Q$).
The resulting $(t, Q)$ pairs serve as inputs for the subsequent event reconstruction.

A simple yet robust Continuous Over-Threshold Integral (COTI) algorithm was selected as the baseline for this data analysis.
COTI identifies pulses by first calculating the waveform baseline and then scanning 8 ns windows for consecutive points above or below a threshold to determine the pulse start and end times, respectively.
The pulse charge is obtained by numerically integrating the waveform between these times.
The pulse time is determined by fitting a straight line to the pulse’s rising edge and finding where it intersects the baseline.
The COTI threshold corresponds to approximately 0.25 p.e.

{\bf Point-like event reconstruction:} Event reconstruction primarily consists of vertex determination, relying mostly on PMT timing; and energy reconstruction, heavily dependent on PMT charge.
After correcting for photon time-of-flight and the event start time, residual hit-time distributions—usually based on the earliest hits—are constructed. 
The corresponding PDFs are primarily shaped by the scintillation time profile and the PMT TTS.
The amount of charge seen by PMT depends on the event energy but also on its position within the LS volume.
Several reconstruction algorithms were used for this analysis and all provided overall consistent results.

OMILREC~\cite{Huang:2022zum}, a combined time- and charge-likelihood method, was developed to simultaneously reconstruct the event vertex and energy. 
Its time likelihood calculation utilizes residual hit-time PDFs, derived from the $^{68}$Ge source at detector center and constructed separately for dynode and MCP PMTs.
The charge likelihood is based on the expected PMT charge response, parameterized using the $^{68}$Ge source at various positions and natural sources like $^{214}$Po.
By integrating both charge and timing information, the method significantly improved vertex reconstruction by about 25\% to 50\%, particularly for events near the detector edge.

In an alternative approach, OMILREC reconstructs the energy relying on a vertex provided by the JVertex algorithm. 
The timing PDFs for JVertex are derived from AmC calibration data obtained with the ACU system and constructed using only dynode PMTs, which offer superior timing performance.
Separate PDFs are produced according to the detected photon multiplicity, to account for cases where multiple photons are registered by the same PMT within its time resolution.

VTREP is an independent algorithm reconstructing both vertex and energy. 
The vertex is reconstructed by identifying the position at which the time-of-flight correction produces the sharpest PMT hit-time distribution, taking into account the detector geometry and light refraction at the acrylic boundary~\cite{Liu:2018fpq}. 
The particle energy is obtained by maximizing a
likelihood function that incorporates the probabilities of both fired and unfired PMTs, based on the predicted light intensity at each PMT.
The light intensity prediction and probability density functions were obtained from uniformly distributed $^{214}$Po events from the $^{214}$Bi-$^{214}$Po decay cascade and laser calibration samples, respectively~\cite{Takenaka:2025hgi}.

\textbf{Muon track reconstruction:} A cluster-finding algorithm is applied to the PMT charge distribution of muon candidates to identify localized regions of high charge associated with muon entry and exit points.
For events with two or more clusters, muon tracks are reconstructed as straight lines connecting the charge-weighted centers of the entry and exit clusters.
The resulting angular distributions of reconstructed muons agree well with MC predictions~\cite{JUNOComiss}.

\subsection * {Detector response model}

In this section we discuss the relationship between the detector response and reconstructed variables and different particle types depositing energy in the detector volume. In an LS detector such as JUNO, this encompasses characterisation of the effective light yield, non-linearity of the energy scale, energy resolution, and the residual spatial non-uniformity across the large detector volume. These effects must be precisely modeled and calibrated to achieve accurate prediction of the energy spectra of reactor neutrino signal as well as all backgrounds.

{\bf Effective light yield:} The energy scale is anchored to the 2.223\,MeV gamma peak originating from neutron capture on protons. 
The effective light yield of approximately 1785 p.e./MeV was calibrated using the neutron-capture peak from an AmC neutron source deployed at the detector center~\cite{JUNOComiss}. 
It encompasses all detector effects, including the number of emitted photons, their propagation through the detector, and the detection probability at the PMTs.

{\bf Energy scale non-linearity:} The visible energy in the LS is not proportional to the deposited energy. Moreover, different particles depositing the same energy produce different numbers of photons. Such non-linearities arise from ionization quenching, dependent on the particle stopping power $dE/dx$, and from the Cherenkov photons. Instrumental nonlinearity may also arise during the charge reconstruction of PMT waveforms.

\begin{figure*}[t]
\centering
\includegraphics[width=0.8\textwidth]{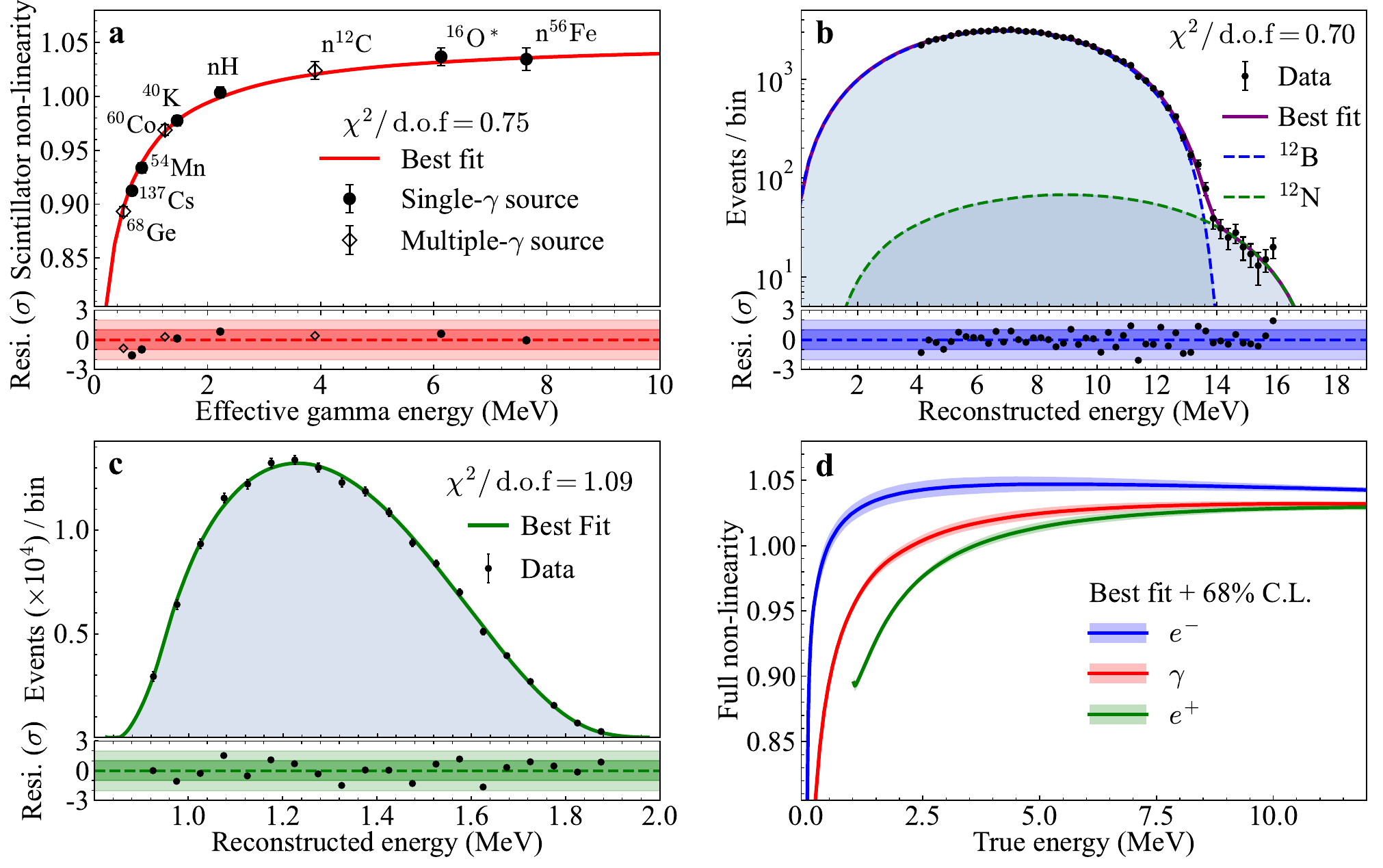}
\caption{{\bf Energy scale non-linearity calibration.} {\bf a}, Scintillator non-linearity from $\gamma$ calibration sources deployed at the detector's center: single-$\gamma$ sources (solid circle), multi-$\gamma$ sources (hollow rhombus), and the best-fit curve (red). {\bf b}, Measured  cosmogenic $^{12}$B $\beta^{-}$ spectrum (points) compared to prediction (blue line), with a best-fit $^{12}$N component (green dashed line)of 2.7\% identified via its high-energy shoulder. {\bf c}, Measured $^{11}$C $\beta^{+}$ spectrum (points) and the best-fit model (green line). {\bf d}, Non-linearity response model for electrons (blue), $\gamma$'s (red), and positrons (green).}
\label{fig:NL}
\end{figure*}

The non-linearity model is constrained using three calibration datasets: (1) reconstructed visible energy from eight $\gamma$ sources with energies ranging from 0.511\,MeV to 6.13\,MeV ($^{137}$Cs, $^{54}$Mn, $^{40}$K, $^{68}$Ge, nH, $^{60}$Co, n$^{12}$C, and $^{16}$O$^{\ast}$), deployed at the detector center. The de-excitation gamma from neutron absorption on Fe (n$^{56}$Fe) in the source capsule was used for additional validation; and two kinds of continuous spectra from high-purity samples of cosmogenic isotopes reconstructed in the FV, namely (2) $^{12}$B electron spectrum  ($Q = 13.37$\,MeV) and (3) $^{11}$C positron spectrum ($Q = 0.96$\,MeV), with visible energy increased by 2\,$\times$\,511\,keV annihilation gammas.

In the fits of $\gamma$ sources energy spectra, individual decay schemes were taken into account.
Pile-up with $^{14}$C was found to having a negligible effect on the peak positions.
%
The model uses the energy spectra of secondary electrons from Compton scattering of $\gamma$'s, obtained from Geant4 simulations, and their stopping power, taken from the ESTAR~\cite{ESTAR} database. Systematic effects including energy leakage from source encapsulation and peak stability were also evaluated. The $^{12}$B sample achieves $>$96\% purity, with the expected $^{12}$N component ($\sim$2.5\%) explicitly included in spectral fit.
Ionization quenching is modelled  via the Birks' law~\cite{Birks:1964zz} with the coefficient $k_B$ being a free fit parameter. Cherenkov light contribution is modeled using either Geant4-based parametric functions or the Frank–Tamm formula~\cite{Frank:1937fk}, with the relative strength controlled by a free parameter $k_C$. An additional parameter $A$ accounts for the absolute scintillation energy scale. The instrumental non-linearity was modeled using either a linear or exponential function, with both forms anchored at the nH capture peak.
All model parameters are determined through $\chi^2$ minimization to the combined calibration data. The best-fit results to the mono-energetic peaks from gamma calibration sources and the continuous energy spectra from cosmogenic isotopes, as well as the predicted electron and positron non-linearities, are shown in Fig.~\ref{fig:NL}. An overall non-linearity uncertainty of 1\% is achieved for positron. 

Three analysis groups adopted slightly different approaches in constructing non-linearity models and usage of calibration data. Comparisons show that including or excluding instrumental non-linearity for centrally located gamma calibration sources has minimal impact; a similar conclusion holds when incorporating n$^{56}$Fe $\gamma$ data. Consequently, the three analyses agree within 0.5\%.

{\bf Energy resolution:} The energy reconstruction resolution for positrons is crucial for the NMO analysis~\cite{JUNO:2024jaw}, but is less critical for determining the solar oscillation parameters presented here. Studies of the energy resolution achieved for $\gamma$'s at the detector center~\cite{JUNOComiss} show slightly worse performance than predicted by MC simulations~\cite{JUNO:2024fdc}.
Resolution for the $^{68}$Ge source is approximately 3.5\%.

\begin{figure*}[!h]
\centering
\includegraphics[width=\textwidth]{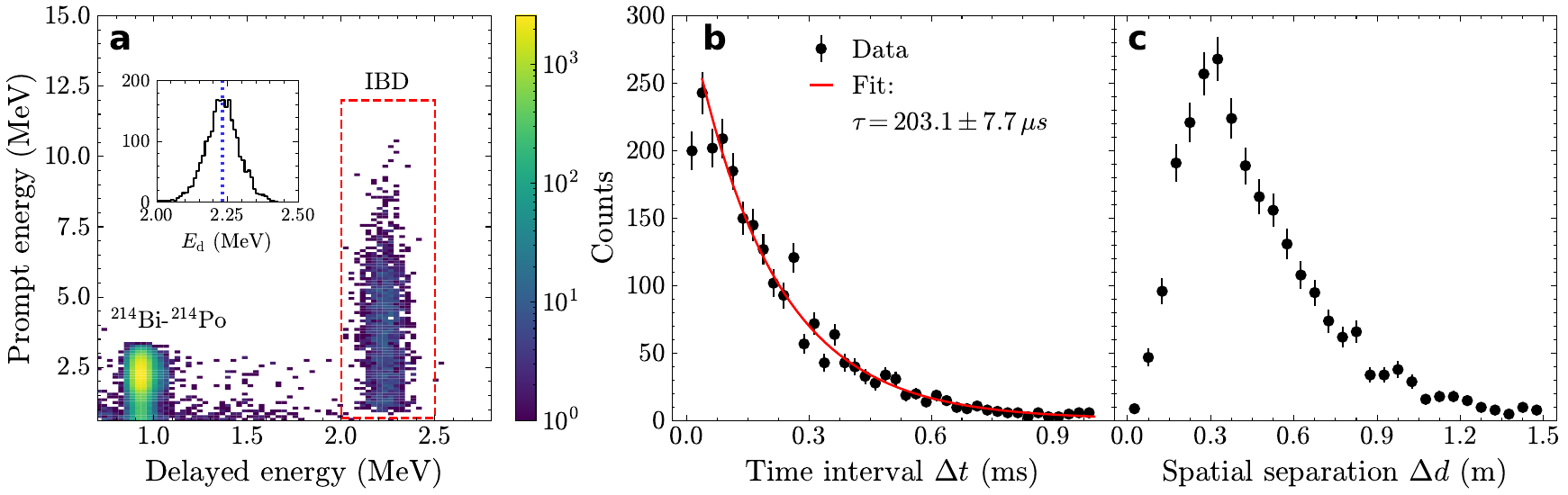}
\caption{{\bf Reactor neutrino candidates characteristics.} {\bf a,} Two-dimensional prompt-versus-delayed energies (with enlarged delayed energy window) showing the $^{214}\mathrm{Bi}$-$^{214}\mathrm{Po}$ cluster at $E_{d}$$\sim$1~MeV and the IBD selection region (dashed red box). The inset shows the delayed-energy spectrum, 
featuring a clear neutron-capture peak at $E_{d}\simeq2.23$~MeV 
(blue dashed line).  {\bf b,} Temporal correlation following an exponential decay with a fitted neutron-capture time of $\tau$=
203.1$\pm$7.7~$\mu$s. The deviation of the first point is due to the $\Delta t >$ 5~$\mu$s cut and is  therefore not included in the fit. {\bf c,} Spatial separation $\Delta d$ of the prompt and delayed vertices.}
\label{fig:ibdCandidates}
\end{figure*}

\subsection * {Antineutrino event selection}
\label{subsec:evtSel}

The selection of reactor neutrino candidates is based on following criteria: 
(1) spatio-temporal prompt-delayed coincidence to maintain high signal efficiency, 
(2) a FV cut to suppress backgrounds caused by external radioactivity (accidental coincidences and fast neutrons from undetected muons), 
(3) dedicated muon veto strategy to reject cosmogenic backgrounds, 
(4) a multiplicity cut to further reject neutron-related background from undetected muons and instrumental backgrounds.
Prior to these selections, the spontaneous light emission in PMT (“flashers”) was suppressed by applying cuts on the standard deviations of the PMT hit multiplicity and timing distributions. The overall detection efficiency was determined through a detailed evaluation of individual selection criteria. Independent analyses using alternative reconstructions and selection criteria confirmed statistically consistent results.

{\bf Prompt-delayed coincidence cuts} applied in the selection are as follows: a prompt energy $0.7 < E_p < 12.0$\,MeV, a delayed energy $2.0 < E_d < 2.5$\,MeV, a time interval $5\,\mu\rm{s} < \Delta t < 1\,\rm{ms}$, and a spatial separation $\Delta d < 1.5$\,m between prompt and delayed signals. 

Energy-selection efficiencies were validated with calibration data. The prompt-energy cut was fully efficient above threshold, as verified using a $^{68}$Ge source. The delayed-energy window retained $99.99\%$ of events from neutron captures on protons with $0.1\%$ relative uncertainty due to energy resolution. The temporal-coincidence efficiency, determined from an exponential fit to the $\Delta t$ distribution of IBD candidates, was $96.8\%$ with a $0.02\%$ relative uncertainty due to capture-time variation. The spatial-correlation cut preserved $98.3\%$ of events with a relative uncertainty of 0.1\%, derived from the reconstructed IBD $\Delta r$ distribution. Both the temporal- and spatial-coincidence efficiencies are verified using AmC sources and MC simulations.

\begin{figure}[t]
\centering
\includegraphics[width=0.48\textwidth]{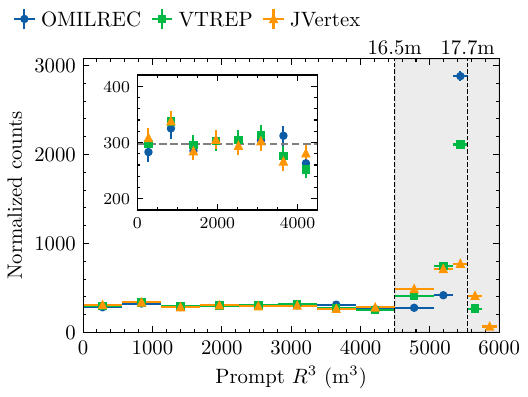}
\caption{{\bf Spatial distribution of reactor neutrino candidates.} Prompt-event counts versus reconstructed cubic radius ($R^3$) are shown for the three vertex algorithms: OMILREC (blue), VTREP (green), and JVertex (orange). Horizontal bars indicate $R^3$ bin widths; vertical dashed lines mark the FV boundary (16.5\,m) and detector edge (17.7\,m). The inset shows the uniform event density within the FV, with the dashed line indicating the mean count level. The rising counts beyond the FV are primarily due to accidental coincidences, with minor contributions from fast- and double-neutron backgrounds.}
\label{fig:fiducial}
\end{figure}

{\bf Fiducial volume cut:} Prompt positron candidates are required to have radius $R<16.5$\,m and $|z|<15.5$\,m  to strongly suppress accidental and cosmogenic neutron backgrounds and to minimize the impact of biases in the reconstruction algorithms. Alternative analyses used different vertex reconstruction algorithms and adopted slightly different criteria, such as applying a separate $R<17.2$\,m for the delayed signal, or 
limiting the polar region cut only around $z$ axis ($|z|>15.5$\,m and $\rho=\sqrt{x^2+y^2}<2$\,m). Across all reconstruction methods and FV definitions, the total efficiency variation remained below 1\%. Figure~\ref{fig:fiducial} shows the radial distributions of prompt IBD candidates using three independent vertex reconstruction algorithms. As it can be seen, the  distributions are uniform within the FV and the rate increase at larger radii is limited to the volume with $R > 16.5$\,m.

The FV efficiency is obtained from the geometrical volume fraction accepted in the selection (80.6\%), under the assumption of a uniform IBD distribution within the FV.
We performed independent estimations using a $^{12}$B sample and a high-energy ($>$3.5\,MeV) IBD sample by calculating the fraction of these events reconstructed inside the FV relative to the total. These results were found to be consistent with the nominal method. 

The spread among the different  reconstruction algorithms near the fiducial boundary yields  a  1.6\% relative FV efficiency uncertainty.  An independent estimate based on the measured $\sim$10 cm vertex-reconstruction bias at the FV boundary, as determined from ACU source deployments, yields a comparable $\sim$1.8\% relative uncertainty. 

{\bf Muon veto}--Events were rejected if the delayed signal occurred within 5\,ms of a preceding muon, 
or if it was found within 4\,m of an identified spallation neutron (SPN) and within 1.2\,s after that neutron (namely spallation-neutron veto). Muon candidates were identified as CD triggers with charge $Q > 3 \times 10^4$ p.e. or WP triggers with $Q > 700$ p.e. The detector-wide veto of 5\,ms following muon events in the CD or WCD results in a livetime efficiency of 95.55\%, with negligible uncertainty. 

Spallation neutrons were identified by their timing ($20\,\mu\text{s} < \Delta t_\mu < 2$\,ms after a muon) and either their charge ($3000 < Q < 3\times 10^4$ p.e.) or, equivalently, their reconstructed energy ($1.5\,<E<\,20~\text{MeV}$). We also tested an additional selection based on the spread of the hit-time distribution to suppress afterpulse contamination and obtained consistent results. 
The spallation-neutron veto acts only in localized regions, and the veto efficiency for these localized events was determined by segmenting the detector into 10-cm cubes and integrating their livetimes, yielding 98.0\% with negligible uncertainty. 
The spallation-neutron veto removes $>$90\% of $^9$Li/$^8$He events.
One independent analysis using a track-based vetoes applied around reconstructed muon trajectories, had a further 25\% suppression of the residual $^9$Li/$^8$He, with a 0.6\% reduction in signal efficiency.

{\bf Multiplicity cut:} It removed events with additional energy deposits above 0.7\,MeV (within the FV) or in the [2.0, 2.5]\,MeV range (outside it) within a [$t_d - 2$\,ms, $t_d + 1$\,ms] window. The multiplicity-cut efficiency evaluated based on the above-mentioned method was 97.4\%. The livetime approach provides a direct, model-independent measurement, while an analytic estimate based on the singles rate serves as a cross-check and agrees with the livetime method. 
Two other analyses using slightly different energy thresholds and fiducial requirements showed $\sim$1\% efficiency variation. 

The characteristics of reactor neutrino candidates are shown in Fig.~\ref{fig:ibdCandidates}. 
The two-dimensional prompt-versus-delayed energy plot shows a clear signal concentration within the selection window (red box), with correlated backgrounds from $^{214}$Bi-$^{214}$Po decay dominating at lower energies. Furthermore, the prompt–delayed signal pairs exhibited strong agreement with predicted temporal and spatial correlations.

\begin{figure}[t]
\centering
\includegraphics[width=0.48\textwidth]{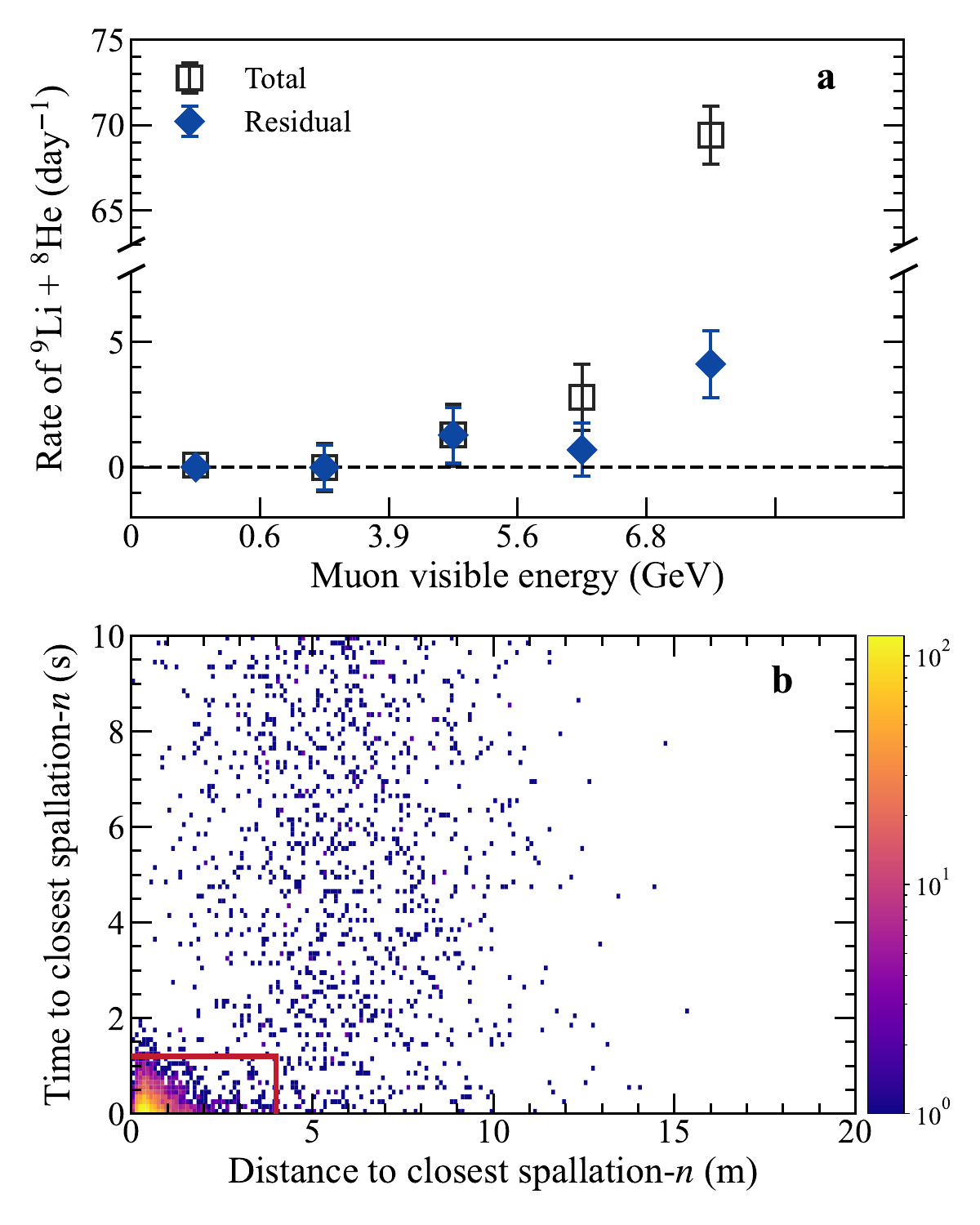}
\caption{{\bf $^9$Li/$^8$He background analysis}. \textbf{a}, Measured $^9$Li/$^8$He rates across muon visible energy, comparing data before and after the SPN veto. \textbf{b}, Spatial- and temporal- distribution of IBD candidates before the SPN cut shown relative to preceding SPN events. A clear cluster of $^9$Li/$^8$He events is visible at the origin. The red box indicates the SPN veto criteria.}
\label{fig:LiHe}
\end{figure}

\subsection * {Backgrounds} 

The following backgrounds remain for IBDs: correlated backgrounds from the $\beta$-$n$ cascade of $^9$Li/$^8$He decays, geoneutrinos, a rare branch in $\beta$-$\alpha$ cascade of $^{214}$Bi-$^{214}$Po decays, remote reactors, fast neutrons, $^{13}$C($\alpha$, n)$^{16}$O, and atmospheric neutrino neutral-current (NC) interactions, and uncorrelated backgrounds from accidental coincidence. The estimated background rates are summarized in Table~\ref{tab:ibd}. Below we provide additional details for some backgrounds.

{\bf Cosmogenic $^9$Li/$^8$He background:} The dominant background originates from cosmogenic $\beta$--$n$ emitters $^9$Li and $^8$He. Their rates were extracted from the time distribution between antineutrino candidates and preceding muons. At high muon rates, however, this method becomes uncertain due to the similar timescales of isotope decays and genuine antineutrino signals. We therefore performed a combined fit across four muon energy intervals, accounting for cross-interval correlations where $^9$Li/$^8$He events from one interval mimic antineutrino signals in others. The $^9$Li/$^8$He ratio was constrained by the measurement from Daya Bay~\cite{DayaBay:2024xye}. Rate uncertainties were obtained from the covariance matrix of the interval counts and verified via a bootstrap method incorporating dataset correlations.

The $^9$Li/$^8$He production is typically accompanied by neutron emission, as indicated by both simulations (Geant4~\cite{GEANT4:2002zbu} and FLUKA~\cite{Battistoni:2015epi,Ahdida:2022gjl}) and previous Daya Bay results~\cite{DayaBay:2012fng}. The spallation-neutron veto mentioned in context rejected over $\sim$90\% of $^9$Li/$^8$He background, exceeding 97\% for muons with visible energy above 10 GeV as shown in Fig.~\ref{fig:LiHe}. One analysis further applied a cylinder veto within 1 m and 0.5 s along muon tracks of single muons without spallation-neutron tags, providing a small additional reduction.

To improve the time-since-last-muon fit, an alternative method selected only muons accompanied by spallation neutrons, which are more likely to produce $^9$Li and $^8$He. This selection reduced the muon rate and alleviated degeneracies in the fit. Using a $^9$Li/$^8$He-enriched sample from IBD candidates within certain distances of muon tracks, the neutron accompaniment efficiency was measured as (90\,$\pm$\,2)\%  by comparing yields with and without the SPN requirement. 
Both methods gave consistent production rates, and the residual $^9$Li/$^8$He background was estimated by subtracting the expected IBD signal. 

A high-purity $^9$Li/$^8$He sample, tagged via associated spallation neutrons, enabled spectral comparison with a spectrum calculated from nuclear databases and an empirical model first developed for Daya Bay~\cite{DayaBay:2016ggj}, constraining spectral shape uncertainty to 20\%. Furthermore, muons passing through the Top Tracker system~\cite{JUNO:2023cbw} provided a calibrated sample to validate the near-exponential lateral distance distribution between $\beta$-n emitters and parent muons.

\textbf{$^{214}$Bi-$^{214}$Po background:} During the LS filling phase, a $^{214}$Bi–$^{214}$Po background was identified. Although the $^{214}$Po $\alpha$-particle energy is strongly quenched and typically falls below the delayed-energy threshold, a subset of higher energy events above the $^{214}$Po peak has been observed.
Such events can originate either from the excited states of $^{214}$Po observed by Borexino~\cite{Borexino:2019gps} or from the $\alpha$-proton elastic scattering, reported by SNO+~\cite{SNO:2025koj}.
In both cases, alpha is accompanied by a particle ($\gamma$ or proton, respectively) with smaller quenching and thus higher visible energy and can eventually extend into the IBD delayed-energy window. This effect 
was quantified using data from the initial high-radon phase of JUNO filling. Scaling the measured tail fraction by the $^{214}$Bi-$^{214}$Po rate during the physics dataset yields a background contribution of (0.18\,$\pm$\,0.10) per day.

\textbf{Accidental background} was measured using an off-time window method, applying identical selection criteria to the standard IBD analysis except for the prompt–delayed coincidence timing. With the FV, the rate was determined to be $(4.9\pm0.3)\times10^{-2}$\,/day. The uncertainty is dominated by off-window event statistics.

\textbf{Atmospheric neutrino neutral-current interactions} on carbon nuclei produce neutrons, generating correlated IBD-like backgrounds~\cite{Cheng:2024uyj}. The rate was estimated using state-of-the-art neutrino interaction models~\cite{Andreopoulos:2009rq,Golan:2012rfa} with the Honda atmospheric flux calculation~\cite{Honda:2015fha}. Model dependence was evaluated by comparing GENIE\,3.0.6~\cite{Andreopoulos:2009rq} and NuWro\,19.02~\cite{Golan:2012rfa} generators, including initial nuclear models and final-state interaction variations, with nuclear de-excitation simulated via TALYS~\cite{Koning:2005ezu}. The model-averaged prediction gives 0.12\,/day (pre-selection) with a conservative 50\% uncertainty covering interaction modeling, de-excitation, and flux systematics. Spectral shape uncertainty was determined from deviations among the selected models.

\textbf{Neutron backgrounds:} Cosmic-ray muons untagged by the WP or passing through the surrounding rock produce energetic neutrons that create correlated backgrounds through two mechanisms: fast neutrons mimic IBD signals via proton recoil followed by neutron capture, while double neutrons arise from sequential capture of two neutrons. The muon veto ($>$99.9\% efficient)  and the stringent FV cut reduced these to negligible levels (see Fig.~\ref{fig:fiducial}) . Simulations predict residual rates of 0.02\,/day (fast neutrons) and 0.05\,/day (double neutrons), concentrated near the LS edge and confirmed with corner-clipping muon data. The fast-neutron spectrum was obtained from muon-tagged events, while the double-neutron prompt spectrum used the IBD delayed-capture shape. Both backgrounds were assigned a conservative 100\% uncertainty, supported by the absence of excess events in the $>12$~MeV IBD sideband (fast neutrons) and near the edge of the FV in Fig.~\ref{fig:fiducial} (double neutrons).

{\bf $^{13}$C($\alpha$, n)$^{16}$O background:} Alpha particles emitted by radioactive contaminants of the LS can be captured on $^{13}$C nuclei (1.1\% isotopic abundance), resulting in the emission of MeV-scale neutrons. A prompt signal can arise before the delayed neutron capture from neutron elastic scattering on protons, its inelastic interactions with $^{12}$C, or from de-excitation $\gamma$'s emitted by $^{16}$O nuclei produced in excited states, yielding a signature that mimics that of an IBD signal. This background has been recently revised using the open-source SaG4n software~\cite{JUNO:2025nbe} as well as Geant4 based simulations. The detector response to reaction products was simulated with JUNO MC. Considering the achieved radiopurity of the LS, this background was found to be very small (0.04\,/day). About 95\% of the backgrounds originates from $\alpha$ decays of $^{210}$Po that is out of secular equilibrium with the $^{238}$U chain. Rate and shape uncertainties of 25\% or 50\% were assigned by different groups considering different approaches in tuning the MC response, in particular to protons and $\alpha$'s.

\begin{figure}[t]
\centering
\includegraphics[width=0.5\textwidth]{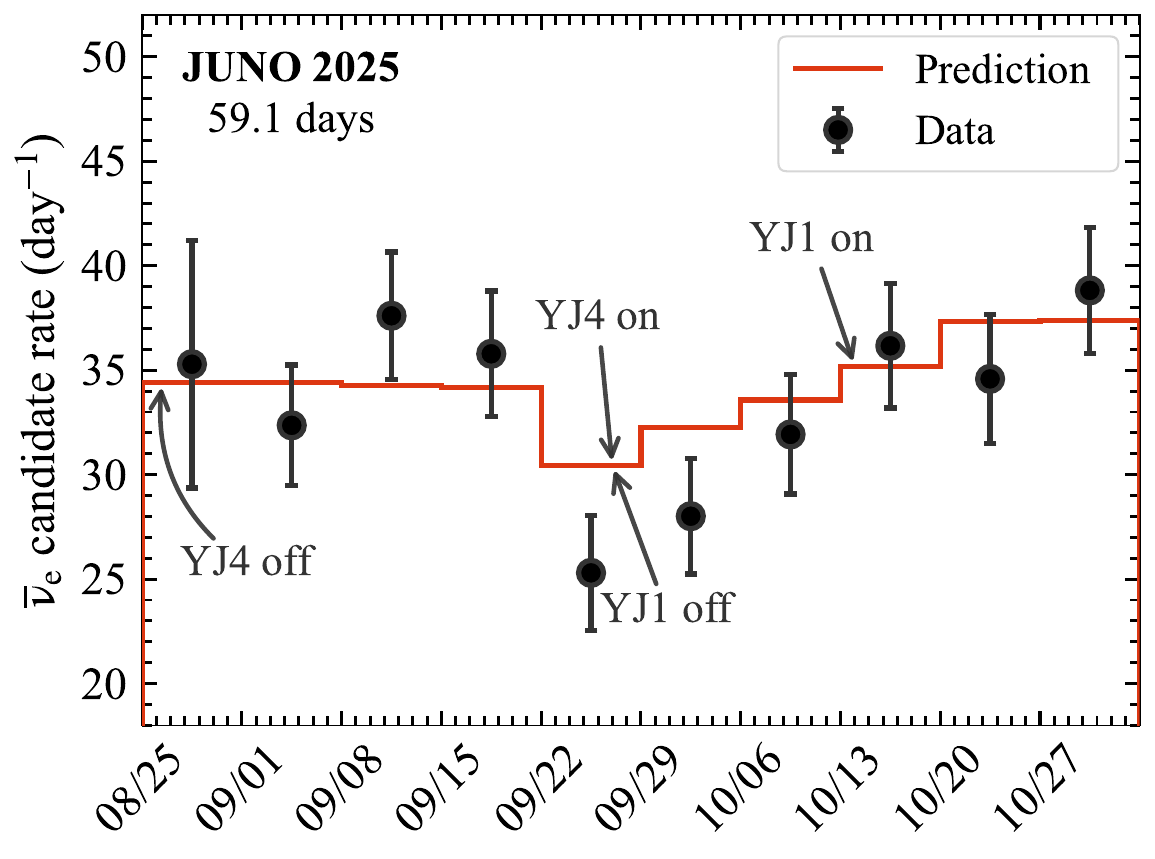}
\caption{{\bf Temporal distribution of reactor neutrinos.} Rates of antineutrino candidates (after subtraction of the mean background rates), shown in one week time bins (black points with statistical error) are compared to the prediction (red line). Arrows indicate operations on the cores YJ1 and YJ4 of the Yangjiang NPP. The NPP reduced the power output due to the occurrence} of Super Typhoon Ragasa on September 24$^{\mathrm{th}}$.
\label{fig:ibdRate}
\end{figure}

\subsection*{Reactor flux prediction and uncertainties}\label{subsec:fluxSpec}

The predicted number of IBD events in reconstructed-energy $i$-th bin is
\begin{align} 
N_i= & C\times\int\limits_{E^\text{rec}_{i}}^{E^\text{rec}_{i+1}} dE^\text{rec} \int\limits_{}^{} dE_{\bar{\nu}_e} \int\limits_{-1}^{1}d\cos\theta \nonumber\\ & \times \sum_{r} \frac {{\cal P}_{ee} (E_{\bar{\nu}_e},L_{r})}{4 \pi L_{r}^{2}} S_{r}(E_{\bar{\nu}_e})\nonumber\\ & \times \frac{d\,\sigma}{d\,\cos\theta}(E_{\bar{\nu}_e},\cos\theta) R(E^\text{rec}|E_{\bar{\nu}_e}),
\label{method:pred} 
\end{align}
where $E^{\rm rec}$ is the reconstructed prompt energy, $E_{\bar\nu_e}$ the antineutrino energy and $L_r$ the baseline from reactor core $r$ to JUNO. 
${\cal P}_{ee}$ is the electron antineutrino survival probability, $S_r$ the emitted spectrum of core $r$, $\theta$ the positron scattering angle, $\frac{d\,\sigma}{d\,\cos\theta}$ the IBD cross section~\cite{Vogel:1999zy}, and $R$ the detector response function. The overall normalization factor $C$ includes the number of free target protons $N_p$, the live time and the IBD selection efficiencies.

$N_p$ was determined from the total LS volume, density and hydrogen mass fraction, using a geometric survey of the acrylic vessel, flow monitoring during filling and dedicated elemental analyses of LS samples. 
Combining these measurements yielded
$N_p = (1.442\,\pm\,0.014)\times 10^{33}$.

For each core $r$, the source spectrum is
\begin{equation}
S_r(E_{\bar\nu_e}) = N_f^r \big[ S_{\rm MI}(E_{\bar\nu_e}) + \Delta S_r(E_{\bar\nu_e}) \big] ,
\end{equation}
where $N_f^r$ is the total number of fissions, derived from the reactor thermal power, fission fractions and energy per fission~\cite{Ma:2012bm}. 
The term $S_{\rm MI}$ is a component common to JUNO and Daya Bay and is parameterised in a segmented, largely model-independent form,
\begin{equation}
S_{\rm MI}(E_{\bar\nu_e}) =
S_0(E_{\bar\nu_e}) + \sum_{i=1}^{N} \alpha_i \, B_i(E_{\bar\nu_e}) \,,
\label{methods:seg}
\end{equation}
where $S_0$ is a smooth reference spectrum, $B_i$ are basis functions localised in neutrino energy $i$-th bin, and the coefficients $\alpha_i$ are free parameters constrained by the Daya Bay measurement. 
This construction allowed the combined JUNO and Daya Bay fit to absorb most uncertainties in the absolute reactor spectrum into $\{\alpha_i\}$, while JUNO remained primarily sensitive to the oscillatory distortions governed by $\Delta m^2_{21}$ and $\sin^2\theta_{12}$.

The residual term $\Delta S_r$ accounts for differences between individual reactors and the Daya Bay configuration, including small changes in effective fission fractions, contributions from spent nuclear fuel (SNF) and non-equilibrium corrections for long-lived fission fragments~\cite{Mueller:2011nm,DayaBay:2016ggj}. 
The SNF contribution was calculated using the refueling history and SNF inventory provided by the NPP, including the shutdown effect~\cite{Ma:2015lsv}.
Reactor thermal power, fission fractions and fission energies were taken from standard reactor simulations~\cite{Ma:2012bm,bib:Appollo_2010,Ma:2014bpa,DayaBay:2016ssb}, with uncertainties at the few-per-mille to percent level. SNF and non-equilibrium corrections are assigned 30\% relative uncertainties but contribute only at the sub-percent level to the total flux. 
Differences in effective fission fractions between the JUNO and Daya Bay data sets were below 1\%, implying that the relative flux normalization between the two experiments is known at $\sim\,0.1\%$ level for a 10\% uncertainty in the isotopic reference spectra.

Uncertainties in thermal power, fission fractions, SNF and non-equilibrium effects were treated as correlated among cores within the same power plant and uncorrelated between different plants. Table~\ref{tab:Uncer} summarizes the uncertainties in the predicted non-oscillated antineutrino signal, which serve as inputs for the calculated $\overline{\nu}_{e}$ rates in Table~\ref{tab:ibd}.
The resulting time evolution of the predicted IBD rate, compared to the measured rate with backgrounds subtracted, is shown in Fig.~\ref{fig:ibdRate}; the observed variations closely track the known reactor power history.

\begin{table}[t]
\begin{tabular}{lc}
\toprule
{\bf Source}  & {\bf Uncertainties} \\
\toprule
Target protons & 1.0\% \\
Reference spectrum & 1.2\% \\
Thermal power & 0.5\% \\
Fission fraction & 0.6\% \\
Spent nuclear fuel & 0.3\% \\
Non-equilibrium & 0.2\% \\
Different fission fraction & 0.1\%\\
\botrule
\end{tabular}
\caption{Summary of detector and reactor related uncertainties on the predicted reactor neutrino rate.  
\label{tab:Uncer}  }
\end{table}

\begin{figure}[!h]
\centering
\includegraphics[width=0.49\textwidth]{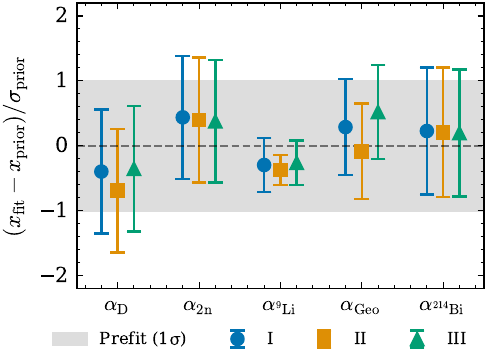}
\caption{{\bf Three analyses post-fit comparison. }
Post-fit pulls of selected nuisance parameters in the three independent JUNO analysis chains. From left to right, these parameters correspond to the detection efficiency, the double-neutron rate, the $^9$Li/$^8$He rate, the geoneutrino rate, and the $^{214}$Bi-$^{214}$Po rate. Coloured markers with vertical error bars show the best-fit shift and $1\sigma$ uncertainty of each parameter relative to its prior, $(x_{\rm fit}-x_{\rm prior})/\sigma_{\rm prior}$, for Analyses~I--III, while the shaded band denotes the $\pm 1\sigma$ prefit range and the dashed line indicates zero pull. }

\label{fig:pulls}
\end{figure}

\subsection*{Statistical methods}\label{subsec:stats}

The first 59.1\,days of JUNO data were analysed with a semi–model-independent strategy that combines JUNO’s prompt-energy spectrum with reactor antineutrino measurements from Daya Bay~\cite{DayaBay:2021dqj,DayaBay:2025ngb,DayaBay2025FullDataRelease} and state-of-the-art reactor flux calculations~\cite{Estienne:2019ujo,Fang:2020emq,Perisse:2023efm}. 
A segmented spectral function (Eq.~(\ref{methods:seg})) was used in the prediction $\mu(\vec{p},\eta,\alpha)$ (Eq.~(\ref{method:pred})) and shared between JUNO and Daya Bay, so that Daya Bay data anchored the overall normalization and coarse spectral shape, while JUNO was primarily sensitive to the fine oscillation pattern. 
Summation-based reactor flux models provided the underlying spectral shape and its theoretical uncertainty at the level of $\mathcal{O}(10$–$20\%)$.

Oscillation parameters were obtained in a frequentist fit using a binned $\chi^2$ function constructed according to the Combined Neyman–Pearson (CNP) method~\cite{Ji:2019yca}:
\begin{multline}
\chi^{2}(\vec{p},\eta,\alpha)
= (\mu - D)^{T} V^{-1} (\mu - D)
+ \lambda^{2}_{\text{nuis}}(\eta) \\
+ \lambda^{2}_{\text{shape}}(\alpha) 
+ \chi^{2}_{\text{osc}}(\sin^{2}\theta_{13},\Delta m^{2}_{31}),
\end{multline}
where $D$ is the observed IBD prompt-energy spectrum and $V$ is the total covariance matrix. 
The vector $\vec{p}$ contains the oscillation parameters of interest, $\eta$ denotes nuisance parameters describing detector response, backgrounds, and overall flux normalization.
The term \(\lambda^2_{\text{shape}}(\alpha)\) incorporated the reactor flux constraints from Daya Bay measurements and their covariance matrix as outlined in Refs.~\cite{DayaBay:2021dqj,DayaBay:2025ngb,DayaBay2025FullDataRelease}.
The CNP prescription defines the bin-wise statistical variance in $V$ in a way that reduces known biases of traditional $\chi^2$ definitions for Poisson data. 
The term $\chi^{2}_{\text{osc}}$ incorporates the published Daya Bay constraints on $\sin^{2}\theta_{13}$ and $\Delta m^{2}_{31}$~\cite{DayaBay:2022orm}, which are used to lift the intrinsic degeneracy in $\Delta m^{2}_{31}$, rather than to fix their values.

The residual dependence on reactor flux modelling was quantified by constructing a nominal summation model based on the latest nuclear databases and generating an ensemble of alternative spectra that encode plausible fine structures. 
These spectra were propagated through ensembles of toy data and analysed with the same fit. 
The additional model-dependent uncertainty on $\Delta m^{2}_{21}$ and $\sin^{2}\theta_{12}$ was below $0.3\%$ after optimizing the segmentation of the spectral function, and was negligible compared with the current total uncertainties.

Three independent analysis chains implemented this framework with different treatments of $V$, spanning statistically dominated to fully correlated systematic descriptions, and yielded mutually consistent results. 
Confidence regions in the $(\sin^{2}\theta_{12},\,\Delta m^{2}_{21})$ plane were obtained from iso-$\Delta\chi^{2}$ contours at 1$\sigma$, 2$\sigma$, and 3$\sigma$ levels. 
The distribution of nuisance-parameter pulls, shown in Fig.~\ref{fig:pulls}, illustrates the stability of the fit and the agreement among the three analyses.

\bibliography{junoOsciNature}

\end{document}